\documentclass[%
aip,
rsi,
amsmath,amssymb,
reprint,
longbibliography
]{revtex4-2}

\usepackage{graphicx}
\usepackage{dcolumn}
\usepackage{bm}
\usepackage[mathlines]{lineno}
\usepackage[utf8]{inputenc}
\usepackage{mathptmx}
\usepackage{multirow}
\usepackage{braket}

\usepackage[english]{babel}

\usepackage{xcolor}
\colorlet{RED}{red}
\colorlet{BLUE}{blue}
\usepackage{dcolumn}
\usepackage{bm}
\usepackage[version=4]{mhchem} 
\usepackage{acronym}
\usepackage{adjustbox}
\usepackage{float}
\usepackage{tikz}
\usepackage{microtype} 
\usepackage{algorithm}
\usepackage{enumitem}

\usepackage{amsmath}
\usepackage{algpseudocode}
\usepackage[caption=false]{subfig}

\DeclareMathOperator*{\argmax}{arg\,max}
\DeclareMathOperator*{\argmin}{arg\,min}
\newcommand{\setb}[1]{\vec{#1}}
\newcommand{\vect}[1]{\vec{#1}}

\usetikzlibrary{calc,shapes.geometric,decorations.pathmorphing,patterns}

\definecolor{background-color}{gray}{0.98}
\usepackage[margin=2.3cm,bmargin=1cm,footnotesep=1cm]{geometry}

\renewcommand{\star}[1]{#1^{*}}
\renewcommand{\Pr}[1]{\text{Pr}\left(#1\right)}
\newcommand{\E}[1]{\mathbb{E}\left[#1\right]}
\newcommand{\normal}[1]{\text{N}\left(#1\right)}
\newcommand{\Var}[1]{\text{Var}\left[#1\right]}
\newcommand{\tb}[1]{\textcolor{black}{#1}}

\begin{document}

\title{Unleashed from Constrained Optimization: Quantum Computing for Quantum Chemistry Employing Generator Coordinate Inspired  Method}
\author{Muqing Zheng}
\affiliation{%
  Pacific Northwest National Laboratory, Richland, Washington, 99354, USA
}
\affiliation{%
  Lehigh University, Bethlehem, Pennsylvania, 18015, USA
}

\author{Bo Peng}
\email{peng398@pnnl.gov}
\affiliation{%
  Pacific Northwest National Laboratory, Richland, Washington, 99354, USA
}

\author{Ang Li}
\affiliation{%
  Pacific Northwest National Laboratory, Richland, Washington, 99354, USA
}
\affiliation{%
  University of Washington, Seattle, Washington, 98195, USA
}

\author{Xiu Yang}
\affiliation{%
  Lehigh University, Bethlehem, Pennsylvania, 18015, USA
}

\author{Karol Kowalski}
\affiliation{%
  Pacific Northwest National Laboratory, Richland, Washington, 99354, USA
 }

\begin{abstract}
Hybrid quantum-classical approaches offer potential solutions to quantum chemistry problems, yet they often manifest as constrained optimization problems. Here, we explore the interconnection between constrained optimization and generalized eigenvalue problems through the Unitary Coupled Cluster (UCC) excitation generators. Inspired by the generator coordinate method, we employ these UCC excitation generators to construct non-orthogonal, overcomplete many-body bases, projecting the system Hamiltonian into an effective Hamiltonian, which bypasses issues such as barren plateaus that heuristic numerical minimizers often encountered in standard variational quantum eigensolver (VQE). Diverging from conventional quantum subspace expansion methods, we introduce an adaptive scheme that robustly constructs the many-body basis sets from a pool of the UCC excitation generators. This scheme supports the development of a hierarchical ADAPT quantum-classical strategy, enabling a balanced interplay between subspace expansion and ansatz optimization to address complex, strongly correlated quantum chemical systems cost-effectively, setting the stage for more advanced quantum simulations in chemistry.
\end{abstract}

\maketitle


\noindent Accurately obtaining ground and excited state energies, along with the corresponding many-body wave functions, is pivotal in comprehending diverse physical phenomena in molecules and materials. This ranges from high-temperature superconductivity in materials like cuprates~\cite{Imada1998metal} and bond-breaking chemical reactions to complex electronic processes in biological and synthetic catalysts with transition metals~\cite{witzke2020bimetallic} or $f$-block atoms~\cite{colin2022ultrahard}. The associated spin, electronic properties, and dynamics are crucial for deciphering the structure-property-function correlation in various fields, including catalysis, sensors, and quantum materials. However, this task becomes exceptionally challenging in the presence of non-trivial quantum effects, such as strong electron correlation, which influence the evolution of nuclei, electrons, and spins under external stimuli. Conventional wave function methodologies, like configuration interaction, coupled cluster, and many-body perturbation theory, are tailored for diverse electron correlation scenarios~\cite{shavitt2009many}. Still, they often fall short in handling complex cases or exhibit prohibitive scaling with increasing system size. Consequently, this has become a vigorous area of computational research, encompassing both classical and burgeoning quantum computing studies. The primary objective is to strike an optimal balance between accuracy and computational scalability.

In the realm of quantum computing, significant strides have been made towards promising near-term hybrid quantum-classical strategies, which includes variational quantum algorithm ~\cite{peruzzo2014variational,mcclean2016theory,romero2018strategies,shen2017quantum,Kandala2017hardware,kandala2018extending,colless2018computation,huggins2020non,Cao2019Quantum,Grimsley2019adaptive,grimsley2019trotterized,verteletskyi2020measurement,mcardle2019variational,mcardle2020quantum,tilly2021variational},
quantum approximate optimization algorithm \cite{farhi2014quantum, stein2022eqc},
quantum annealing \cite{bharti2021noisy,albash2018adiabatic},
Gaussian boson sampling \cite{Aaronson2011},
analog quantum simulation \cite{trabesinger2012quantum,georgescu2014quantum},
iterative quantum assisted eigensolver~\cite{mcardle2019variational,motta2020determining,parrish2019quantum,kyriienko2020quantum},
and many others. These approaches typically delegate certain computational tasks to classical computers, thereby conserving quantum resources in contrast to exclusively to quantum methods. Within this framework, the variational quantum eigensolver (VQE) and its adaptive derivatives are seen as the frontrunners in leveraging near-future quantum advantages~\cite{peruzzo2014variational,Grimsley2019adaptive}. However, these anticipations are also potentially impeded by the heuristic nature inherent in the critical optimization processes. Issues such as the rigor of ansatz exactness, the challenges in navigating potential energy surfaces replete with numerous local minima, and the numerical nuances in minimization techniques, remain nebulous. These queries are further convoluted when considering the scalability of these methods concerning the number of operators or the depth of quantum circuits involved.

As an alternative to VQE, which employs a highly nonlinear parametrization of the wave function, other near-term strategies aim to construct and traverse a subspace within the Hilbert space to closely approximate the desired state and energy. Typical examples include the quantum subspace expansion~\cite{colless2018computation,mcclean2017hybrid,Takeshita2020Increasing,McClean2020Decoding,urbanek2020chemistry}, the hybrid and quantum Lanczos approaches~\cite{Suchsland2021algorithmicerror,motta2020determining,tkachenko2023quantum}, quantum computed moment approaches~\cite{seki2021quantum,kyriienko2020quantum,kowalski2020quantum,Vallury2020quantumcomputed,Aulicino2022State}, and quantum equation-of-motion approach~\cite{PhysRevResearch.2.043140}. These strategies often draw inspiration from truncated configuration interaction approaches or explore the Krylov subspace (see, e.g., Ref.~\citenum{motta2024subspace} for a recent review on subspace methods for electronic structure simulations on quantum computers). However, the comparison and interrelation of these methods, especially when contrasting subspace expansion with nonlinear optimization, frequently remain unclear. This uncertainty is exacerbated by the choice of ans{\"a}tze, which rarely guarantees exactness.

Recently, motivated from the generator coordinate approaches\cite{hill1953nuclear,Griffin1957collective,rodriguez2002correlations,bender2003self,ring2004nuclear,yao2010configuration,egido2016state,hizawa2021generator}, we developed a generator coordinate inspired nonorthogonal quantum eigensolver as an alternative near-term approach~\cite{qugcm}, potentially addressing the limitations identified in the VQE/ADAPT-VQE.
Similar to its contemporaries, the Generator Coordinate Inspired Method (GCIM)  employs low-depth quantum circuits and efficiently utilizes existing ans\"{a}tze to explore a subspace, targeting specific states and energy levels. Nonetheless, the original GCIM approach demands a priori knowledge of the system to meticulously select both the ans{\"a}tze and generators, thereby circumventing heuristic approaches but potentially affecting its scalability and efficiency.

In this study, we present novel quantum-classical hybrid approaches, inspired by the GCIM approach, aimed at establishing a theoretically exact yet automated subspace expansion procedure. This approach addresses the optimization challenges inherent in conventional VQE methods while maintaining an adaptive framework. Recent efforts have either focused on practical solutions to mitigate some of these optimization limitations—such as leveraging domain-specific knowledge from classical quantum chemistry to construct high-quality wave functions~\cite{PRXQuantum.4.030307}—or have moved toward automated circuit-subspace VQE and its adaptive algorithms, though they still require optimization~\cite{hirsbrunner2024diagnosinglocalminimaaccelerating,martidafcik2024spincouplingneedencoding}.In our approach, we use the conventional Unitary Coupled Cluster (UCC) excitation generators as the basis for subspace expansion, establishing a lower bound on the constrained optimization problem typically encountered in VQE. More importantly, we introduce an optimization-free, gradient-based automated basis selection method from the UCC operator pool. This allows us to develop a hierarchical ADAPT quantum-classical strategy that enables controllable interplay between subspace expansion and ansatz optimization. Our preliminary results suggest that these new approaches not only excel in addressing strongly correlated molecular systems but also significantly reduce simulation time, facilitating deployment on real quantum computers.

\section*{Results}
\subsection*{Comparison between VQE and GCIM approaches: general eigen-problem vs. constrained optimization}
\noindent We provided a brief review of generator coordinate methods (GCM) in Supplementary Information (Section I). A primary advantage of GCM is that variation occurs in the generating function, rather than directly on the scalar generator coordinate. This distinction enables the variational problem to be addressed through a straightforward eigenvalue process, contrasting with solving a constrained optimization problem, the advantages of which can even be demonstrated with a simple toy model.

\begin{figure*}[!htbp]
    \includegraphics[width=\linewidth]{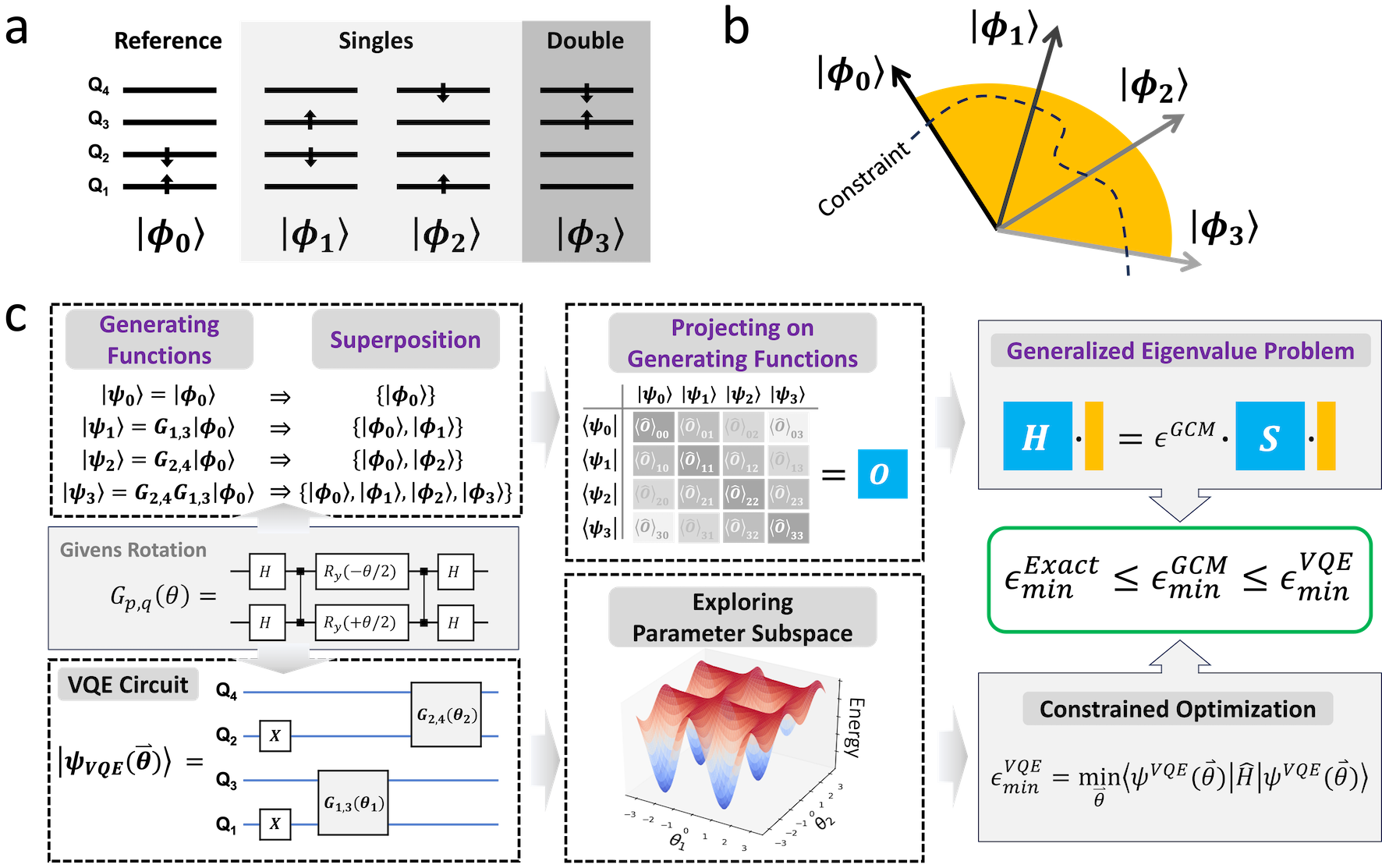}
    \caption{A comparison between GCIM and VQE on a two-electron system. (a) A toy model consists of two-electron (one alpha electron and one beta electron) in four spin-orbitals (two alpha spin-orbitals and two beta spin-orbitals), where the spin-flip transition is assumed forbidden. (b) The projection of the exact wave function on each configuration (yellow shadow). A constrained optimization of the free parameters will put limits on the projection (dashed line).  (c) Comparative demonstration between GCIM method and VQE. The GCIM method generates a set of non-orthogonal bases, called generating functions. Then, the GCIM explores the projection of the system on these generating functions and solves a corresponding generalized eigenvalue problem for the target state and its energy. The conventional VQE essentially explores the parameter subspace for a given wave function ansatz through numerical optimization that can be usually constrained by many factors ranging from ansatz inexactness to barren plateaus and others. For given Givens rotations that generated excited state configurations, the lowest eigenvalue obtained from the GCIM method guarantees a lower bound of the most optimal solution from the VQE. It is worth mentioning that: (i) for a standard (single-circuit) VQE, having one two-qubit Givens rotation acting on a standard quantum-chemistry reference (restricted Hartree-Fock) would not improve the energy estimate; (ii) ``fermionic swap'' gates\cite{PhysRevLett.120.110501} would be required if the excitations included in the ans\"{a}tze involve spin-orbitals that are not mapped onto adjacent qubits for a given quantum architecture.}
    \label{fig:gcm_vqe}
\end{figure*}

Building on this, as depicted in Figure~\ref{fig:gcm_vqe}, we explore the ground state of a toy model comprising two electrons and four spin orbitals using a VQE approach. The VQE approach is mathematically represented as:

\begin{align}
E_{g} &= \min_{\vect{\theta}} \langle  \psi_{VQE}(\vect{\theta}) | H | \psi_{VQE}(\vect{\theta}) \rangle,
\end{align}
where $|\psi_{VQE}(\vect{\theta}) \rangle$ employs Trotterized UCC single-type ans{\"a}tze

\begin{align}
|\psi_{VQE}(\vect{\theta}) \rangle = G_{2,4}(\theta_2) G_{1,3}(\theta_1) |\phi_0\rangle  \label{eq:Givens} 
\end{align}
with  $|\phi_0\rangle $ being a reference state and  $G_{p_i, q_i}(\theta_i) = \exp[\theta_{i}(A_{p_i,q_i} - A_{p_i,q_i}^\dagger)]$ being a single excitation Givens rotation and $A_{p,q} = a_p^\dagger a_q$. The VQE approach here explores a two-dimensional parameter subspace. The consecutive actions of $G_{2,4}$ and $G_{1,3}$ generate four distinct configurations, including the reference. However, due to fewer free parameters than distinct configurations required, the VQE approach is constrained, unable to fully explore the configuration space to guarantee the most optimal solution, irrespective of the numerical minimizer used. In contrast, GCIM solver employs the same Givens rotations, separately or in the product form as discussed by Fukutome~\cite{{fukutome1981group}}, to create four generating functions. These functions correspond to a subspace consisting of four non-orthogonal superpositioned states, providing a scope sufficiently expansive to encapsulate the target state. Here, each Givens rotation is essentially a UCC single circuit, simplifying the quantum implementation compared to its double or higher excitation analogs. 

From the performance difference we observe that 
\begin{itemize}
    \item Although the disentangled UCC ans\"{a}tze can be crafted to be exact~\cite{evangelista2019exact}, the ans\"{a}tze employed in GCIM approach or in the generalized eigenvalue problem need not be exact. The only prerequisite for the ans\"{a}tze used in this process is their ability to generate sufficient superpositioned states for a better approximation to the target state to be found in the corresponding subspace.
    \item The $N$-qubit Givens rotations ($N\ge 2$) used in both VQE and GCIM solver may originate from the same set, ensuring a comparable level of complexity in the quantum circuits involved. To achieve the same level of exactness, VQE would require a deeper circuit by incorporating additional Givens rotations, while the GCIM algorithm necessitates more measurements (than a single VQE iteration). A direct comparison of the total quantum resources utilized between the two methods would hinge on both the number of generating functions required in the GCIM and the number of iterations needed in VQE.
    \item In the context of non-orthogonal quantum eigensolvers,\cite{k-UpCCGSD-paper,PRXQuantum.4.030307,martidafcik2024spincouplingneedencoding,Anselmetti_2021} entangled basis sets are usually employed by applying entanglers on a single determinant reference. Some of these approaches rely on the modified Hadamard test of Ref.~\citenum{huggins2020non} for evaluating the off-diagonal elements at a polynomial cost. Others~\cite{martidafcik2024spincouplingneedencoding,Anselmetti_2021} often require domain-specific knowledge available in classical quantum chemistry to construct high-quality wave functions. Nevertheless, automating the sophisticated ansatz construction process to facilitate easier deployment of the non-orthogonal quantum solver, especially for non-experts, remains a challenge.
\end{itemize}
These observations then lead us to the pivotal question in advancing a more efficient GCIM algorithm for both classical and quantum computation: ``\textit{How can we efficiently select the generating functions to construct a subspace in the Hilbert space that encompasses the target state?}''


\subsection*{The scaling of number of generating functions: From exponential to linear}
\begin{figure*}[!htp]
    \includegraphics[width=0.8\linewidth]{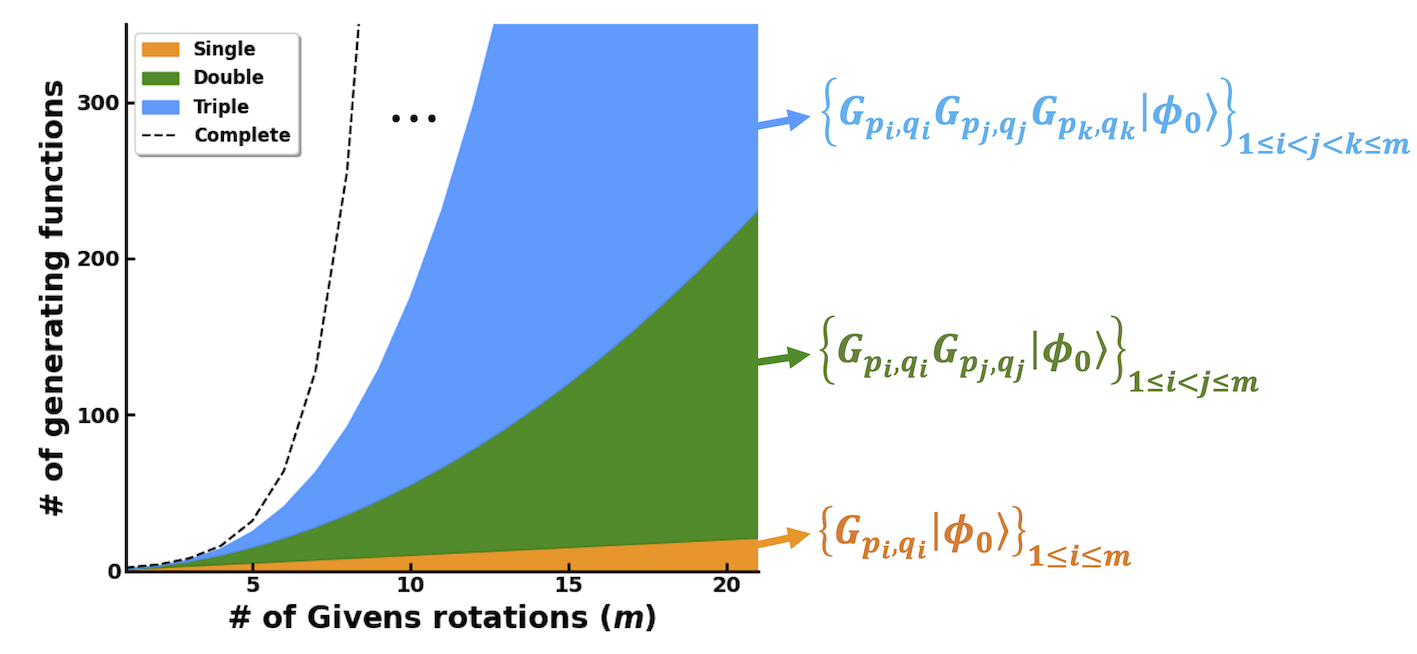}
    \caption{Number of generating functions as functions of the number of Givens rotations (denoted as $m$) included in the GCIM wave function ans\"{a}tze at different truncation levels.}
    \label{fig:num_GenFxn}
\end{figure*}

\noindent In constructing a non-trivial UCC ansatz for a molecular system comprising $n_e$ electrons in $N$ spin orbitals, one might consider applying a sequence of $K \le n_e$ Givens rotations to a reference state $|\phi_0\rangle$ to generate all the possible configurations. For example, if each Givens rotation corresponds to a single excitation, the disentangled UCC Singles ansatz can be expressed as:

\begin{align}
|\psi(\vect{\theta}) \rangle = \prod_{i=1}^K G_{p_i,q_i}(\theta_i) |\phi_0\rangle, \label{eq:ans}
\end{align}
where each $G_{p_i,q_i}$ generates a superposition of no more than two states. Therefore, the total number of configurations, $n_c$, within the superpositioned ans\"{a}tze (\ref{eq:ans})$-$where $\vect{\theta} = \{\theta_i | i=1,\cdots,K\}$ varies$-$cannot exceed $2^K$. This number indicates the maximum number of generating functions utilized to achieve the most optimal solution within the corresponding subspace but exhibits an exponential scaling relative to the count of Givens rotations applied. A strategy to potentially generate the full set of $2^K$ generating functions involves applying $1,2,\cdots,N$ two-qubit Givens rotations separately to the reference, as shown below:
\begin{itemize}[leftmargin=15pt]
    \item $|\phi_0\rangle,$
    \item $G_{p_i,q_i}(\theta_i)|\phi_0\rangle,~~1 \le i \le K,$
    \item $G_{p_i,q_i}(\theta_i)G_{p_j,q_j}(\theta_j)|\phi_0\rangle,~~1 \le i < j \le K,$
    \item $G_{p_i,q_i}(\theta_i)G_{p_j,q_j}(\theta_j)G_{p_k,q_k}(\theta_k)|\phi_0\rangle,~~1 \le i < j < k \le K,$
    \item $~~~~\vdots$
    \item $\prod_{i=1}^K G_{p_i,q_i}(\theta_i)|\phi_0\rangle$
\end{itemize}
Here, the number of generating functions employing $k$ Givens rotations is equivalent to the number of $k$-combinations from the set of $K$ Givens rotations, aligning with the combinatorial identity $\sum_{k=0}^K \bigg( \begin{array}{c} K \\ k \end{array} \bigg) = 2^K$ (see Figure~\ref{fig:num_GenFxn}). Importantly, when $K = n_e$, each sequence of Givens rotations in the above list can be tailored to include a unique excited configuration within the associated subspace, thereby mapping the entire configuration space to ensure exactness in this limit. 
This characteristic bears notable similarity to full configuration interaction treatments, with the distinctive aspect that this approach constructs a non-orthogonal many-body basis set. In practice, to capture strong correlation with a fewer number of generating functions, generalized $N$-qubit Givens rotations are commonly applied in quantum chemical simulations. A typical example includes the particle-hole one- and two-body unitary operators frequently employed in VQE and ADAPT-VQE simulations. Notably, it has been demonstrated that products of only these particle-hole one- and two-body unitary operators can approximate any state with arbitrary precision.\cite{evangelista2019exact}

It's important to recognize that the generating functions, including the ans\"{a}tze, are not generally orthogonal. This non-orthogonality necessitates extra resources for evaluating the associated off-diagonal matrix elements and the entire overlap matrix for solving a generalized eigenvalue problem. Additionally, if higher-order Givens rotations are involved, the configuration subspace expansion will be influenced by the ordering of the Givens rotations in the ans\"{a}tze. Nevertheless, the ordering limitation can be partially circumvented by allowing the possible permutations of the Givens rotations within the generating functions.

The strategy outlined above can also be employed to establish a hierarchy of approximations to the GCIM wave function, targeting a configuration subspace where the ansatz (\ref{eq:ans}) is fully projected out. For example, the approximation at level $k$ ($0\le k < K$) can be defined by a working subspace that incorporates both the ansatz (\ref{eq:ans}) and some generating functions produced by acting a product of at most $k$ Givens rotations on the reference. Notably, when $k=K$, regardless of the rotations in the generating functions, the configuration subspace has expanded sufficiently for the ansatz (\ref{eq:ans}) to be fully projected out, completely removing the constraints in the optimization. For other approximations where $0\le k<K$, a variational procedure can be undertaken before the configurational subspace expansion, aiming to deduce a lower bound for the optimal expectation value of $H$ with respect to the ansatz (\ref{eq:ans}). In this fashion, the GCIM solver can be implemented as a subsequent step either after each VQE iteration or following a complete VQE calculation to unleash some constraints due to the local minima and the inexactness of the ansatz. 


\subsection*{Integration of GCIM with ADAPT approach}
\noindent To facilitate a more robust selection of generating functions, we consider the fluctuation of the GCIM energy when a new generating function is included. In the context of VQE, similar considerations have led to the development of its ADAPT (Adaptive Derivative-Assembled Pseudo-Trotter) version. In the GCIM framework, several routines can be proposed for a gradient-based ADAPT-GCIM approach. A straightforward method is to directly compute the GCIM energy gradient with respect to the scalar rotation, as detailed in the Supplementary Information (Section III). The energy gradient computed in this way depends on the eigenvector solved from the GCIM general eigenvalue problem after the inclusion of new generating functions. However, when solving a generalized eigenvalue problem, the eigenvalue (i.e. the energy) is more sensitive to the working subspace than to the scalar rotation in the employed generating functions. This sensitivity is exemplified in the \textbf{Lemma 1} of \textbf{Theorem 1} in the Supplementary Information (Section II), where, for a $2\times 2$ generalized eigenvalue problem, the choice of the scalar rotations in the employed generating functions is relatively flexible. 

\begin{figure*}[!htbp]
    \includegraphics[width=0.8\linewidth]{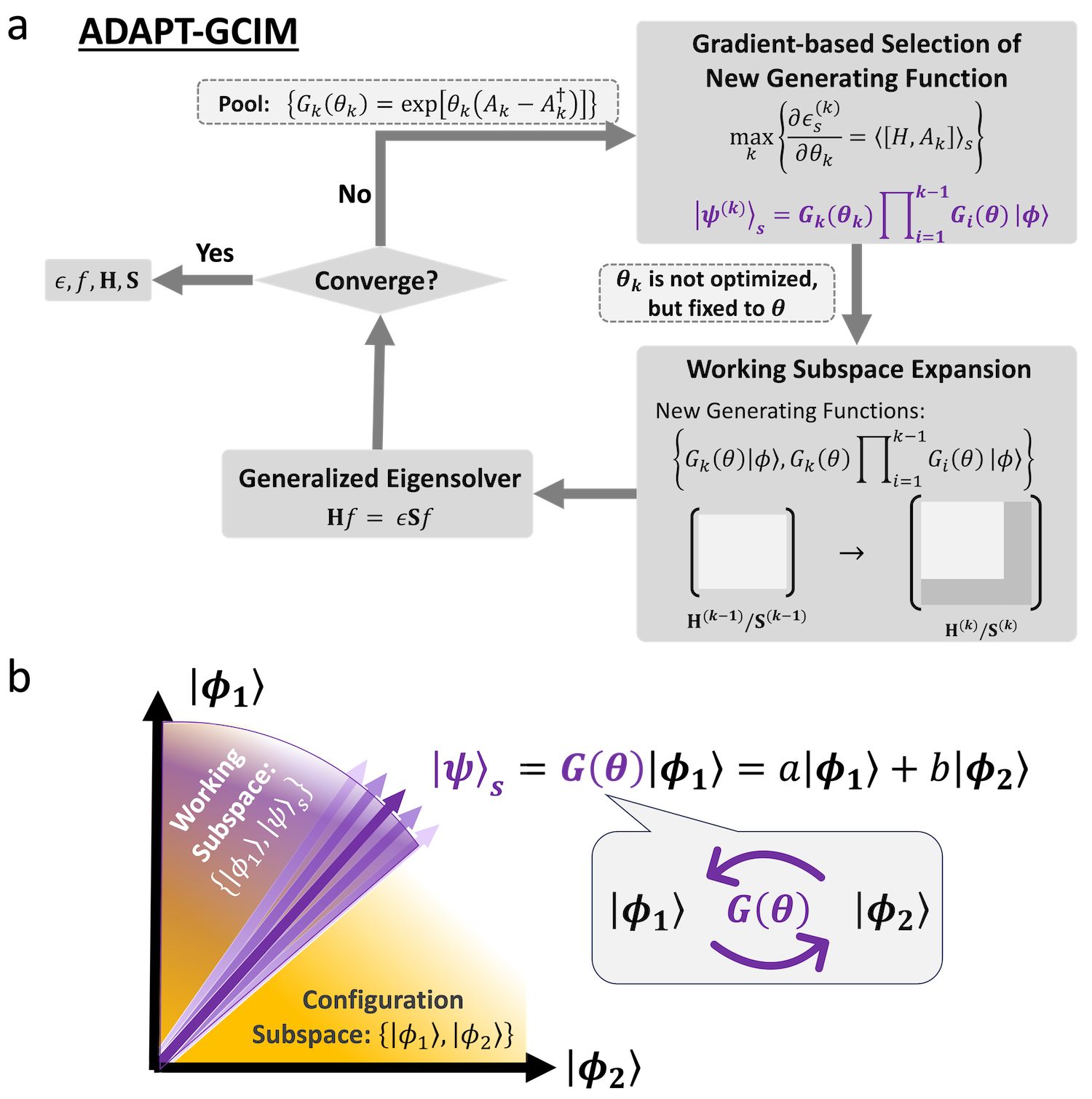}
    \caption{Overview of ADAPT-GCIM and the effects of Givens rotations. (a) Schematic depiction of the ADAPT-GCIM algorithm in $(n+1)$-th iteration. The working subspace and its corresponding configuration subspace are characterized by a surrogate state $|\psi\rangle_s$. (b) In a two-configuration subspace, a Givens rotation can couple two configurations, effectively generating a surrogate state, a superposition of two configurations, that can be employed as a metric to characterize this subspace. The quality of the working subspace, comprising one configuration and the surrogate state, is not sensitive to the specific choice of scalar rotation.}
    \label{fig:adapt_gcm}
\end{figure*}

This consideration leads us to propose a more flexible, gradient-based ADAPT-GCIM algorithm (illustrated in Figure~\ref{fig:adapt_gcm}a). In this approach, we use a surrogate product state $|\psi\rangle_s$ that, irrespective of its specific non-singular scalar rotations, primarily serves as a metric to (i) abstract the change in the GCIM working subspace and (ii) approximate the GCIM energy gradient calculation. Utilizing this metric, we can compute the gradient in a manner similar to that in the conventional ADAPT-VQE. However, a significant difference from ADAPT-VQE is that in ADAPT-GCIM, we skip the optimization of phases, which as we will see later significantly reduces the simulation time.  Also, the energy in ADAPT-GCIM is not directly obtained from the measurement of the expectation value with respect to the surrogate state. Instead, it is obtained from solving a generalized eigenvalue problem within the working subspace.


\subsection*{Numerical and Real Hardware Experiments}
\begin{figure*}[!htbp]
    \includegraphics[width=\linewidth]{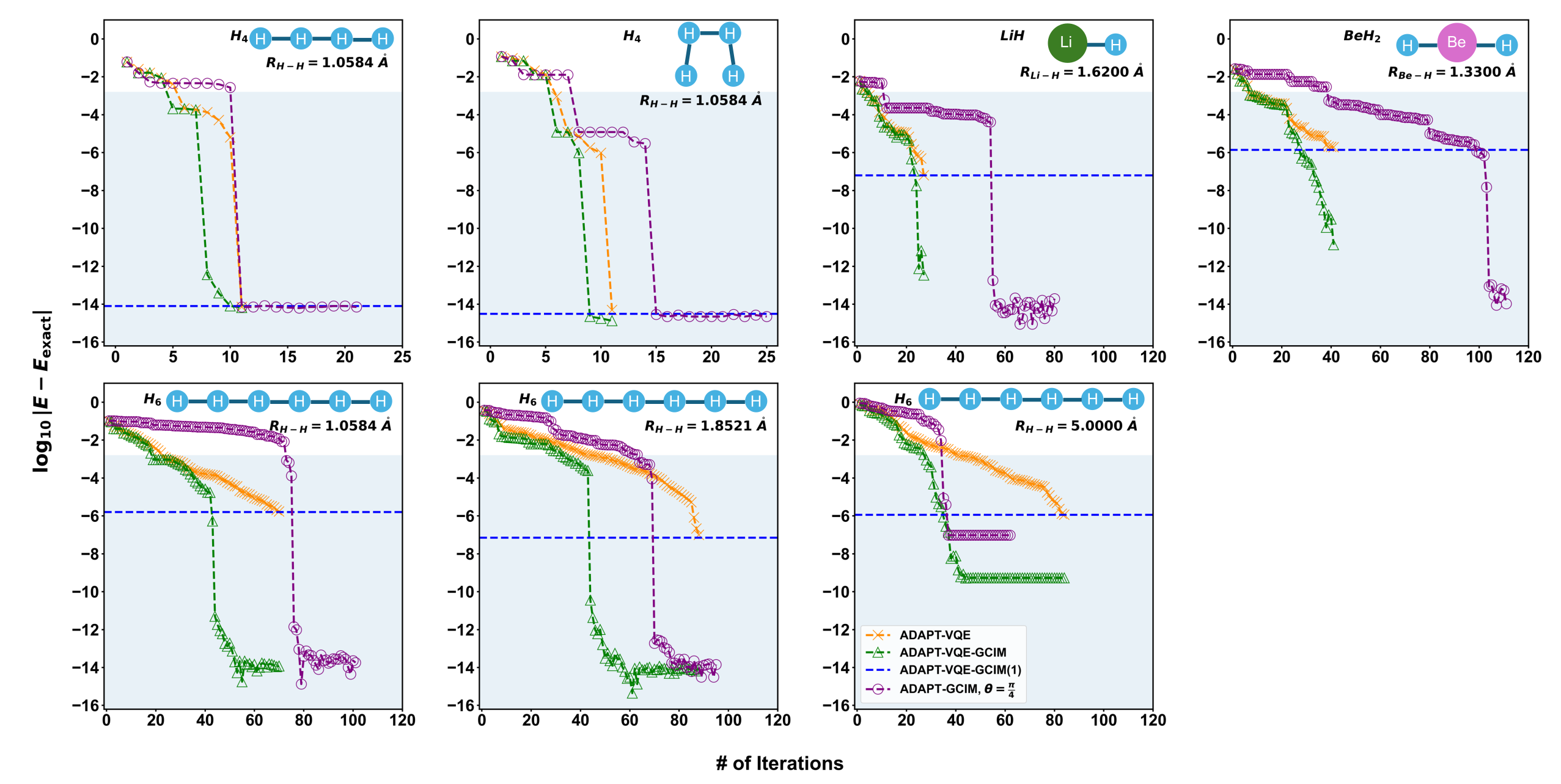}
    \caption{Convergence performance of ADAPT-GCIM in computing the ground states of four molecules. For comparison, results from ADAPT-VQE, ADAPT-VQE-GCIM, and ADAPT-VQE-GCIM(1) are also included. The shaded region is when error is smaller than the chemical accuracy. ADAPT-VQE and ADAPT-VQE-GCIM are set to converged when the sum of the magnitudes of all gradients falls below $10^{-4}$. ADAPT-GCIM is terminated if the changes on lowest eigenvalue are under $10^{-6}$ a.u. for at most 10 (for $H_4)$ or 25 (the other molecules) consecutive iterations, where this threshold depends on the size of the operator pool and the amount of unchosen ans{\"a}tze.}
    \label{fig:example-comp}
\end{figure*}

\noindent We have performed numerical experiments on four molecular systems across seven geometries to test the performance of our proposed ADAPT-GCIM approach in searching the ground states corresponding to different electronic configurations. All experiments use STO-3G basis. For instance, the more pronounced quasidegeneracy in the almost-square $H_4$ compared to linear $H_4$~\cite{qugcm}, and in stretched $H_6$ as opposed to compact $H_6$, presents strong static correlations that pose a challenge for traditional single-reference methods~\cite{jankowski1980applicability}. Additionally we proposed two GCIM modified ADAPT-VQE approaches: ADAPT-VQE-GCIM and ADAPT-VQE-GCIM(1) for performance analysis. In the ADAPT-VQE-GCIM approach, a GCIM step is performed after each ADAPT-VQE iteration, while in the ADAPT-VQE-GCIM(1) approach, a `one-shot' GCIM calculation is performed at the conclusion of an ADAPT-VQE calculation (see Supplementary Information (Section IV) for more details). The performance results, as shown in Figure~\ref{fig:example-comp}, include comparisons with ADAPT-VQE. It is evident that the GCIMs offer several strategies for locating a lower bound to the ADAPT-VQE energy. In all the test cases, the ADAPT-VQE-GCIM and ADAPT-GCIM approaches exhibit steady monotonic converging curves, almost always reaching close to machine precision. 

To verify the accuracy, we reconstruct the ADAPT-GCIM approximation to the ground-state vector of $H$ by
\begin{align}
    \ket{\psi_{GCIM}} = \mathcal{N} \cdot \sum_j f_j \ket{\psi_j} , 
\end{align}
where $f$ is the eigenvector solved from the generalized eigenvalue problem in ADAPT-GCIM at the final iteration and $\{\ket{\psi_j}\}$ is the selected basis set. 
$\mathcal{N}$denotes the normalization factor. Table~\ref{tab:exact-overlap} provides the overlaps between the exact ground-state vector, $\ket{\psi_{exact}}$, and corresponding GCIM approximation for all molecules in this work, numerically proving the accuracy of the ADAPT-GCIM method.
Furthermore, compared to ADAPT-VQE, GCIM approaches perform better in strongly correlated cases. For example, in the case of H$_6$, as the H$-$H bond length increases, the number of ADAPT-GCIM or ADAPT-VQE-GCIM iterations decreases, while the number of ADAPT-VQE iterations increases.

\begin{table}
    \centering
    \caption{Overlaps between ADAPT-GCIM and exact ground-state vectors (machine precision is $2.22 \times 10^{-16}$).}
    \label{tab:exact-overlap}
    \begin{tabular}{cc} \hline \hline
     Molecule            & $1 - |\braket{\psi_{GCIM} | \psi_{exact} }|^2$   \\ \hline   
    $H_4$ (linear)      & $2.22 \times 10^{-16}$ \\ [0.2cm]
    $H_4$ (square)      & $<2.22 \times 10^{-16}$\\ [0.2cm]
    $LiH$            & $2.22 \times 10^{-16}$\\ [0.2cm]
    $BeH_2$           &   $2.22 \times 10^{-15}$\\ [0.2cm]
    $H_6$ ($1.0584\,\,\mathring{A}$)  & $1.55 \times 10^{-15}$\\ [0.2cm]
    $H_6$ ($1.8521\,\,\mathring{A}$)  & $2.66 \times 10^{-15}$\\ [0.2cm]
    $H_6$ ($5.0000 \,\, \mathring{A} $)  & $2.42 \times 10^{-2}$ \\ \hline \hline
    \end{tabular}
\end{table}

Using the same set of circuits, the GCIM formalism, in addition to ground-state energies, provides estimates of the excited-state energies corresponding to low-lying excited states. In Table ~\ref{tab:exec-energy}, we compare excitation energies corresponding to EOMCCSD (equation-of-motion CC approaches with singles and doubles) \cite{eomccsd}, EOMCCSDT (equation-of-motion CC approaches with singles, doubles, and triples) \cite{eomccsdt}, and the ADAPT-GCIM approaches. The character of excited states collated in Table~\ref{tab:exec-energy} correspond to states of mixed configurational character dominated by single and double excitations from the ground-state Hartree-Fock determinant (SD) and challenging states dominated by double excitations (D). For the challenging low-lying doubly excited state of the $H_6$ system at $R_{H-H}=1.8521$ $\mathring{A}$, the ADAPT-GCIM approach yields excitation energy in good agreement with the EOMCCSDT one and outperforming the accuracy the EOMCCSD estimate.

\begin{table}
    \centering
    \caption{Excitation energies in eV from different methods}
    \label{tab:exec-energy}
    \begin{tabular}{ccccc} \hline \hline
  \multirow{2}{*}{Molecule}             &Character of       &EOMCC-        &EOMCC-          &ADAPT-          \\
              &excited state  &SD   &SDT  &GCIM \\ \hline
    $H_4$ (linear)      &SD(1 excited)   &12.738     &12.555    &12.565   \\ [0.2cm]
    $H_4$ (square)      &D (1 excited)   &4.275      &4.273     &4.183  \\ [0.2cm]
    $LiH$            &SD (1 excited)   &3.588      &3.586     &3.586  \\ [0.2cm]
    $BeH_2$           &SD (3 excited)  &15.558     &14.563    &14.545  \\ [0.2cm]
    $H_6$ ($1.0584\,\,\mathring{A}$)       &SD (1 excited)   &9.846      &9.386     &9.477  \\ [0.2cm]
    $H_6$ ($1.8521\,\,\mathring{A}$)       &D (1 excited)   &2.927      &1.581     &1.340 \\ \hline \hline
    \end{tabular}
\end{table}


\begin{figure*}[!htbp]
    \includegraphics[width=\linewidth]{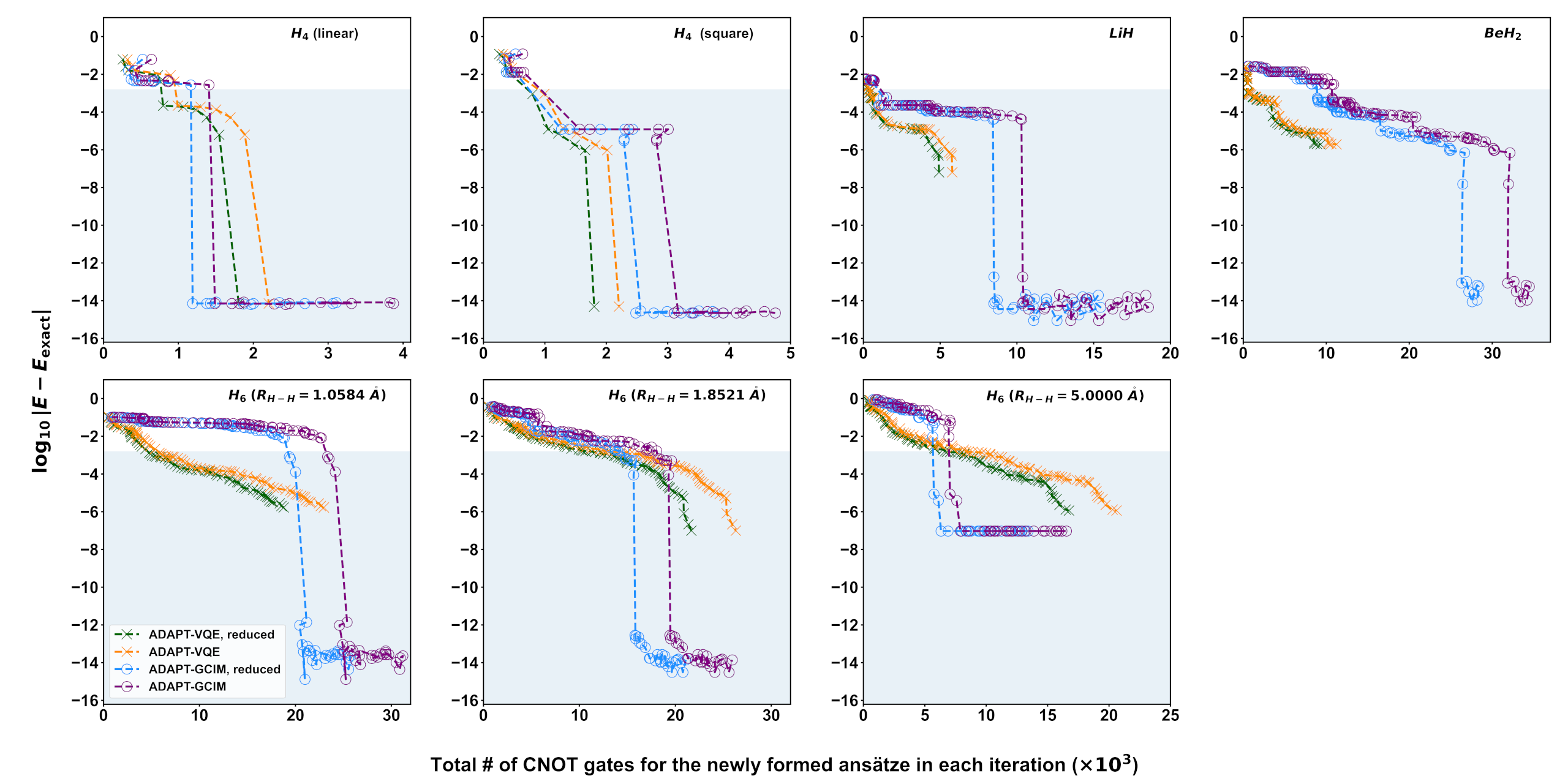}
    \caption{Energy difference as a function of the number of CNOT gates in the circuits of newly formed ans{\"a}tze. For ADAPT-VQE, such ansatz is the product of operators that prepare the VQE state $\ket{\psi_{VQE}}$. For ADAPT-GCIM, the newly formed ans{\"a}tze are the newly selected Givens rotation operator and the product of Givens rotation operators. All circuits are generated from the order-1 trottered fermionic operators by Qiskit. The reduced counts are obtained using \textit{qiskit.transpile} with optimization level 3, which includes canceling back-to-back CNOT gates, commutative cancellation, and unitary synthesis. Our estimations to ADAPT-VQE match with the corresponding calculations in Ref.~\citenum{tang2021qubit}. A further discussion on the calculation of the number of CNOT gates is in the Supplementary Information (Section V).}
    \label{fig:cnot-count}
\end{figure*}

In assessing the quantum resources needed for the ADAPT approaches, it is apparent that ADAPT-VQE-GCIM and ADAPT-VQE-GCIM(1) require additional resources compared to ADAPT-VQE approach to improve energy results. On the other hand, the direct comparison between ADAPT-VQE and ADAPT-GCIM is not straightforward. For example, the number of CNOT gates in ADAPT-VQE and ADAPT-GCIM simulations depends on the ansatz preparation and the number of measurements to reach a desired accuracy. Figure~\ref{fig:cnot-count} demonstrates the total number of CNOTs for preparing the following ans\"{a}tze: 
\begin{itemize}
    \item $\prod_{i = 1}^{k} G_{i}(\theta_i)|\phi_{\rm HF}\rangle$ at the $k$-th ADAPT-VQE iteration;
    \item $G_k(\theta_k)|\phi_{\rm HF}\rangle$ and $\prod_{i = 1}^{k} G_{i}(\theta_i)|\phi_{\rm HF}\rangle$ at the $k$-th ADAPT-GCIM iteration,
\end{itemize}
at each energy error level. Here, $G_{i}$ is either UCC single or double excitation operator, and chosen from the same operator pool. As can be seen, the ADAPT-GCIM, compared to ADAPT-VQE, appears requiring fewer CNOT gates for more strongly correlated cases.

Regarding measurements in fault tolerant quantum computation, ADAPT-VQE requires measurements for VQE and gradient calculations at each iteration. The number of VQE measurements depends on the average optimizations per iteration ($\tilde{N}_{\rm opt}$) and the total number of iterations ($N_{\rm iter}$). Gradient measurements are roughly equal to the size of the pool multiplied by the number of unitary terms in the Hamiltonian ($N_{\rm term}$), which can be reduced with grouping techniques. Thus, total measurements for ADAPT-VQE scales as $\mathcal{O}(\tilde{N}_{\rm opt}\times N_{\rm iter} + N_{\rm term}\times N_{\rm iter})$. In ADAPT-GCIM, measurements are needed for gradients and constructing the $\mathbf{H}$ and $\mathbf{S}$ matrices. The latter scales as the squared number of generating functions, $\mathcal{O}(N_{\rm GF}^2)$, but per iteration, it only grows as $\mathcal{O}(N_{\rm GF})$, assuming a linear increase of generating functions with iterations. Since $N_{\rm GF}\sim \mathcal{O}(N_{\rm iter})$, measurement comparison between the two ADAPT approaches reduces to $\mathcal{O}(\tilde{N}_{\rm opt}\times N_{\rm iter (VQE)})$ vs. $\mathcal{O}(N_{\rm iter (GCIM)}^2)$. 

It is worth mentioning, for weakly or moderately correlated cases, ADAPT-GCIM usually requires more iterations than ADAPT-VQE as shown in Figure~\ref{fig:example-comp}. The reason could be partially attributed to the ordering of the operators due to the skip of the parameter optimization, as well as the non-orthogonal nature of the bases. To enforce orthogonality of the GCIM bases, a secondary subspace projection can be performed on the effective Hamiltonian as illustrated in Section VI of the Supplementary Information, where the numerical simulations indeed exhibit slightly better convergence performance from some molecules (see LiH and H$_6$ performance curves in Figure S1). To improve the convergence due to the ordering of the operators, in Section VIII of the Supplementary Information, we consider a middle ground between ADAPT-VQE-GCIM and ADAPT-GCIM by allowing intermittent and truncated optimization, where for every $m$ ADAPT iterations (``intermittent''), at most $n$ rounds (``truncated'') of classical optimization can be conducted on ansatz parameters over the corresponding objective function. The proposed ADAPT-GCIM($m,n$) approach indeed provides a significant improvement on the convergence speed: ADAPT-GCIM(5,2) reaches $10^{-6}$ error level faster than ADAPT-GCM and ADAPT-VQE as evidenced in Figure S3 for two H$_6$ test cases. Also, as shown in Table~\ref{tab:bigtime}, the extra optimization rounds and simulation time brought by introducing the intermittent and truncated optimization to the ADAPT-GCM can be very minimal.

\begin{table*}
\caption{Simulation costs for energy calculations of seven geometries using the ADAPT approaches on a laptop equipped with an Apple M3 Max chip. During the optimization rounds, the default optimizer is the Broyden–Fletcher–Goldfarb–Shanno (BFGS) algorithm, implemented in SciPy~\cite{scipy}. ADAPT-GCIM(5,2) is an extended method introduced in the Supplementary Information (Section VIII).}\label{tab:bigtime}
\begin{tabular}{ccccccc} \hline \hline
\multirow{2}{*}{Molecule} & \multirow{2}{*}{ADAPT-} & {Min \# of ADAPT} & {Total \# of Optim-} & \multicolumn{2}{c}{Total Simulation Time (s)}  & Ground-State \\ \cline{5-6} 
                          &                         & Iterations        & ization Rounds                    & Gradients        & Energy Eval.      &  Energy Error (in a.u.)\\ \hline
\multirow{2}{*}{$H_4$ (linear)}  & \tb{GCIM} & \tb{11}   &0      & \tb{0.08}     & \tb{0.38}    &\tb{7.11 $\times 10^{-15}$}   \\
                      & VQE & 11    &121     & 0.10     & 7.61    &7.11 $\times 10^{-15}$   \\ [0.2cm]
\multirow{2}{*}{$H_4$ (square)}  & \tb{GCIM} & \tb{15}    &0     & \tb{0.11}     & \tb{0.73}  &\tb{2.89 $\times 10^{-15}$}    \\
                      & VQE & 11    &136     & 0.09     & 17.44     &4.88 $\times 10^{-15}$ \\ [0.2cm]
\multirow{2}{*}{$LiH$}  & \tb{GCIM} & \tb{56}    &0     & \tb{2.53}     & \tb{30.50}    &\tb{8.88 $\times 10^{-15}$}  \\
                      & VQE & 27    &439     & 1.24     & 253.64   &6.36 $\times 10^{-8}$  \\ [0.2cm]
\multirow{2}{*}{$BeH_2$} & \tb{GCIM} & \tb{104}   &0     & \tb{10.05}    & \tb{242.14}    &\tb{8.70 $\times 10^{-14}$} \\
                      & VQE & 41   &1248      & 3.69     & 2308.14  &1.91 $\times 10^{-06}$  \\ [0.2cm]
\multirow{3}{*}{$H_6$ ($1.0584\,\,\mathring{A}$)} & \tb{GCIM} & \tb{79}   &0     & \tb{3.77}     & \tb{77.55}  &\tb{1.33 $\times 10^{-15}$}   \\
                      & VQE & 70    &4792     & 3.17     & 4627.37  &1.70 $\times 10^{-6}$ \\ 
                      &\tb{GCIM(5,2)}  & \tb{67}  &\tb{28}   &\tb{3.35}    & \tb{85.47}  &\tb{6.66 $\times 10^{-15}$} \\ [0.2cm]
\multirow{2}{*}{$H_6$ ($1.8521\,\,\mathring{A}$)} & \tb{GCIM} & \tb{70}   &0     & \tb{3.42}    & \tb{55.85}  &\tb{1.90 $\times 10^{-13}$} \\
                      & VQE & 88   &9209      & 3.99     & 11924.84   &9.98 $\times 10^{-08}$\\ [0.2cm]
\multirow{3}{*}{$H_6$ ($5.0000\,\,\mathring{A}$)} & \tb{GCIM} & \tb{37}   &0     & \tb{1.79}     & \tb{12.03}  & \tb{9.57 $\times 10^{-8}$}  \\
                      & VQE & 84   &11313      & 3.86     & 12765.16  & 1.15 $\times 10^{-6}$ \\ 
                      &\tb{GCIM(5,2)}  & \tb{26}  &\tb{12}  &\tb{1.24}  &\tb{13.41}  &\tb{9.61 $\times 10^{-8}$} \\ \hline \hline
\end{tabular}
\end{table*}

To account for the impact of optimization on the performance difference, we list the total rounds of classical optimization and simulation times for the energy calculations of seven geometries using the ADAPT approaches on the same computing platform in Table~\ref{tab:bigtime}. Remarkably, even though the circuit for obtaining the $\mathbf{H}$ and $\mathbf{S}$ matrices in ADAPT-GCIM is different from the Hadamard circuit used in ADAPT-VQE, where the off-diagonal matrix elements requiring more CNOT gates for simultaneous base preparation, the ADAPT-GCIM approaches still significantly outperform the ADAPT-VQE in simulation time, in particular for more strongly correlated cases. For example, for the strongly correlated H$_6$ molecule at a stretched H$-$H bond length of $5.0000~\mathring{A}$, the GCIM approaches completed in 12$\sim$13 seconds, while ADAPT-VQE required approximately four hours to achieve a similar level of accuracy. It is important to note that stretched molecules often exhibit multi-reference characteristics, posing significant challenges for single-reference methods. However, quantum approaches, including VQEs and GCIM, with appropriately designed circuits/operators that account for this multi-reference nature should, in principle, offer superior simulation performance.

In practice, real quantum measurements often encounter significant noise, especially in solving the generalized eigenvalue problem where finite shot uncertainty is a concern~\cite{qugcm}. Stronger correlations typically require substantially more measurements than weaker ones. As detailed in the Supplementary Information (Section VII), errors in the $\mathbf{S}$ matrix have a greater impact than those in the $\mathbf{H}$ matrix. However, with importance sampling, accuracy can still be improved by orders of magnitudes in most cases despite these challenges (see Figure~S2). We deployed the ADAPT-GCIM computation of the strongly correlated linear H$_4$ molecule on an IBM superconducting quantum computer, $ibm\_osaka$, to conduct a rudimentary test of the applicability of the GCIM approaches on emerging hardware. As shown in the Supplementary Information (Section IX), with problem-specific error mitigation methods, the error in ground-state energy estimation was reduced from 0.046 a.u. to $3.9\times 10^{-9}$ a.u.


\section*{Discussion}
\noindent The GCIM approach stands out among other subspace expansion and generalized eigenvalue problem-solving methods due to its size-extensiveness, attributed to its use of UCC ans\"{a}tze as bases. The approach's scalability for larger systems will be enhanced by its direct application to qubit space, using the exponential of anti-hermitian Pauli strings as generating functions. This strategy aligns with developments like qubit-ADAPT-VQE, which offers scalability for the ADAPT-VQE method~\cite{tang2021qubit}.

The straightforwardness of the GCIM ans\"{a}tze makes it a promising solver also for excited state computation and Hamiltonian downfolding, as demonstrated in the ADAPT-GCIM procedure (Figure~\ref{fig:adapt_gcm}a). 
This lays the foundation for advanced quantum simulations in chemistry. 
To further characterize the quality of the ADAPT-GCIM wave function, we find the UCC-type ansatz that maximizes the overlap with the ADAPT-GCIM optimized state and find the corresponding UCC energy.
That is, we need to find $\ket{\psi_{UCC}^*} := \ket{\psi_{UCC}(\vect{\theta}^*)}$ where
\begin{align}
    \vect{\theta}^* := \argmax_{\vect{\theta}} \,\, |\braket{\psi_{GCIM} | \psi_{UCC}(\vect{\theta})}|^2.
\end{align}
The most straightforward scheme is to use the UCCSD ansatz. We illustrate the quality of the UCCSD ansatz defined in such a way  through eight examples in Table~\ref{tab:qflow-translator}, where $E_{exact}$ is the exact ground-state energy and $E_{UCCSD}$ is its approximation from $\ket{\psi_{UCCSD}^*}$. The optimizer for UCCSD ansatz is the limited-memory BFGS (L-BFGS) algorithm.

\begin{table}
    \centering
    \caption{Overlap between GCIM state and the optimized UCCSD states, and the ground-state energy error of the UCCSD states.}
    \label{tab:qflow-translator}
    \begin{tabular}{ccc} \hline \hline
     \multirow{2}{*}{Molecule}            & \multirow{2}{*}{$1 - |\braket{\psi_{GCIM} | \psi_{UCCSD}^* }|^2$}  &$|E_{UCCSD} - E_{exact}|$ \\ 
     & & (in a.u.) \\ \hline  
    $H_4$ (linear, $0.7500\,\,\mathring{A}$)      & $2.53\times 10^{-6}$ & $ 9.81 \times 10^{-6}$\\ [0.2cm]
    $H_4$ (linear, $0.8000\,\,\mathring{A}$)      & $4.18\times 10^{-6}$ & $ 1.46 \times 10^{-5}$ \\ [0.2cm]
    $H_4$ (linear, $1.0584\,\,\mathring{A}$)      & $6.07 \times 10^{-5}$ & $ 1.29 \times 10^{-4}$ \\ [0.2cm]
    $H_4$ (square, $1.0584\,\,\mathring{A}$)      & $4.43\times 10^{-4}$ & $ 9.81 \times 10^{-4}$\\ [0.2cm]
    $LiH$            &  $3.76\times 10^{-6}$ & $ 1.10 \times 10^{-5}$\\ [0.2cm]
    $BeH_2$   & $2.89\times 10^{-4}$     &  $4.06\times 10^{-4}$\\ [0.2cm]
    $H_6$ ($1.0584\,\,\mathring{A}$)  &  $4.90\times 10^{-4}$ & $ 8.26 \times 10^{-4}$ \\  [0.2cm]
    \tb{$H_6$ ($1.8521\,\,\mathring{A}$)} & \tb{$1.71 \times 10^{-2}$}  &\tb{$1.19\times 10^{-2}$ } \\
    \hline \hline
    \end{tabular}
\end{table}

Last but not least, while the gate counts in Figure~\ref{fig:cnot-count} match with the estimations for fermionic operators in Ref.~\citenum{tang2021qubit}, the CNOT counts can be greatly reduced by the implementations in other works in the future~\cite{berry2007efficient, Anselmetti_2021, arrazola2022universal, kottmann2023molecular}. Another important direction to explore is to evade Hadamard test for evaluating off-diagonal matrix entries in ADAPT-GCIM to effectively reduce the hardware requirements in the circuit implementation\cite{huggins2020non, cortes2022quantum, motta2024subspace}.

\section*{Methods}

\noindent The operator pool employed in all the ADAPT simulations in the present study consists of generalized singly and doubly spin-adapted excitation operators, as explicitly shown in the Supplementary Information (Section V). In all the ADAPT-VQE simulations, the convergence is achieved when the norm of the gradient vector is less than $10^{-4}$. Here, we use a more strict criteria than the one in Ref.~\citenum{Grimsley2019} to show the full picture of the performance of ADAPT-VQE. Other criteria, such as the magnitude of the change of the lowest eigenvalue, can be used for earlier convergence with the assistance of ADAPT-VQE-GCIM. 

For ADAPT-GCIM, the algorithm terminates when the change in the lowest eigenvalue is smaller than $10^{-6}$ a.u. after $T = \min\{T_{auto}, T_{usr}\}$ number of consecutive ADAPT iterations, where $T_{auto}$ is set to $20\%$ of the number of unselected operators in the pool and $T_{usr}$ is a user-defined constant. In this work, $T_{usr}$ is set to $10$ for $H_4$ and $25$ for the other molecules, reflecting different sizes of operator pools. The reason why the termination condition is heuristic, rather than gradient-based like ADAPT-VQE, is that ADAPT-GCIM uses VQE's objective function, as an approximation, along with unoptimized parameter values during the ansatz selection. Hence, at the later iterations, while ADAPT-GCIM may already provide highly accurate energy estimations, the norm of the gradient vector at the operator selection can still remain relatively large.

One issue in the GCIM simulations is the numerical instability of solving the generalized eigenvalue problem. It is possible that the overlap matrix $\mathbf{S}$ becomes numerically indefinite at certain iterations. In classical computing, such issue can be perfectly avoided by orthogonalizing the generating function basis set. However, in quantum computing, the quantum version of the orthogonalization algorithm like the quantum Gram-Schmidt process usually requires extra quantum resources, such as QRAM~\cite{Zhang2021}. 
In the present study, when a singular overlap matrix is encountered, we choose to project both $\mathbf{H}$ and $\mathbf{S}$ using the eigenvectors of $\mathbf{S}$ corresponding to large positive eigenvalues~\cite{urbanek2020chemistry}. Without the disturbance due to the finite number of shots, the threshold for a sufficiently ``large'' eigenvalue can be as relaxed as $10^{-13}$. Otherwise, under the finite-sampling noise (see Section VII of the Supplementary Information), this threshold is usually set to $10^{-5}$ and $10^{-6}$. A comparison test in the Supplementary Information (Section VI) shows that this mitigation is very effective in achieving the accuracy levels reachable by GCIM methods.

In the proposed ADAPT-VQE-GCIM(1), the number of bases in the working subspace equals to the number of Givens rotations in the ansatz generated at the last ADAPT-VQE iteration, incremented by one. This increment accounts for the product of these Givens rotations corresponding to the ansatz itself. Conversely, in the ADAPT-GCIM and ADAPT-VQE-GCIM approaches, the number of bases is twice the number of iterations. To illustrate this, consider the $k$-th iteration ($k>1$): when the Givens rotation $G_k$ with the greatest gradient is selected, two associated basis vectors, 
\begin{itemize}[leftmargin=15pt]
    \item $G_k(\theta_k)|\phi_{\rm HF}\rangle$ and
    \item $G_k(\theta_k)\prod_{i=1}^{k-1} G_i(\theta_i)|\phi_{\rm HF}\rangle$
\end{itemize}
are added to the working subspace. When $k=1$, the basis set is $\{|\phi_{\rm HF}\rangle, G_1(\theta_1)|\phi_{\rm HF}\rangle\}$. So the number of bases is increased by two in every iteration. The key difference is that in ADAPT-GCIM, the phase is set to a constant scalar (e.g., $\pi/4$) for any Givens rotation in the ansatz, while in ADAPT-VQE-GCIM, the phases $\theta_1$ to $\theta_{k}$ are optimized during the $k$-th iteration. In all the experiments of ADAPT-GCIM, $\theta_k$ is set to $\pi/4$ for all iterations.

The implementation of ADAPT-VQE uses the original code in Ref.~\citenum{Grimsley2019} from the corresponding repository~\cite{adaptvqe-repo}, and its computation is based on the SciPy sparse linear algebra package~\cite{scipy}. To have the same foundation for comparison, calculations in Figure~\ref{fig:example-comp} for three GCIM algorithms follow the same manner. The quantum resources estimation, shown in Figure~\ref{fig:cnot-count}, utilizes Qiskit~\cite{Qiskit} for trotterization and circuit generation for excitation operators.  
The data in Table~\ref{tab:qflow-translator} were computed on NERSC Perlmutter supercomputer.
The rest of numerical experiments were carried on a laptop with an Apple M3 Max chip.

\section*{Data Availability}

\noindent The datasets generated and/or analysed during the current study are available in the GitHub repository,
\href{https://github.com/pnnl/QuGCM}{(https://github.com/pnnl/QuGCM)}.

\section*{Code availability}

\noindent The underlying code [and training/validation datasets] for this study is available in ``QuGCM'' GitHub repository and can be accessed via this link \href{https://github.com/pnnl/QuGCM}{https://github.com/pnnl/QuGCM}.

\section*{Acknowledgements}

\noindent M.~Z., B.~P. and K.~K. acknowledge the support by the ``Embedding QC into Many-body Frameworks for Strongly Correlated  Molecular and Materials Systems'' project, which is funded by the U.~S. Department of Energy, Office of Science, Office of Basic Energy Sciences (BES), the Division of Chemical Sciences, Geosciences, and Biosciences under FWP 72689. B.~P. also acknowledges the support from the Early Career Research Program by the U.~S. Department of Energy, Office of Science, under FWP 83466. A.~L. and K.~K. also acknowledges the support from Quantum Science Center (QSC), a National Quantum Information Science Research Center of the U.S. Department of Energy (under FWP  76213). X. Y. was supported by National Science Foundation CAREER DMS-2143915. M. Z. and X. Y. both also were supported by Defense Advanced Research Projects Agency as part of the project W911NF2010022: {\em The Quantum Computing Revolution and Optimization: Challenges and Opportunities}. This research used resources of the Oak Ridge Leadership Computing Facility, which is a DOE Office of Science User Facility supported under Contract DE-AC05-00OR22725. This research used resources of the National Energy Research Scientific Computing Center (NERSC), a U.S. Department of Energy Office of Science User Facility located at Lawrence Berkeley National Laboratory, operated under Contract No. DE-AC02-05CH11231. The Pacific Northwest National Laboratory is operated by Battelle for the U.S. Department of Energy under Contract DE-AC05-76RL01830.\\

\section*{Author Contributions}

\noindent B.~P. conceptualized the idea and designed the research,  M.~Z. and B.~P. implemented the approaches and performed the simulations. All authors analyzed the data and contributed to paper writing.\\

\section*{Competing interests}

\noindent All authors declare no financial or non-financial competing interests. 


\appendix


\section{Generator Coordinate: A Brief Overview} \label{sec:GCM_review}

Originating from studies of nuclear structure and nuclear reactions, including those occurring in heavy-ion collisions or astrophysical scenarios, the Generator Coordinate Method (GCM), proposed by Griffin-Hill-Wheeler (GHW)~\cite{hill1953nuclear,Griffin1957collective}, approximates the many-body wave function $|\Psi\rangle$ of the Hamiltonian $H$ by describing collective motion through the variation of one or more collective generator coordinates, denoted as ${\setb{\alpha}} = \{\alpha_i\}$, where $i=1,\cdots,N_\alpha$ and $N_\alpha>1$. This variation leads to the representation of the many-body wave function of a quantum system in terms of a set of basis states (also called generating functions) $\{|\phi(\setb{\alpha})\rangle\}$:

\begin{align}
    |\Psi\rangle \approx \int {\rm d}{\setb{\alpha}} |\phi({\setb{\alpha}})\rangle f({\setb{\alpha}}). \label{gcm_wfn}
\end{align}
Here, $f({\setb{\alpha}})$ is an unknown ${\setb{\alpha}}$-dependent weight function. The choice of these generator coordinates ${\setb{\alpha}}$ depends on the specific system under study. The fundamental concept of the GCM is to treat these generator coordinates as variational parameters, which are then used to approximate the target wave function as a superposition of $\{|\phi({\setb{\alpha}})\rangle\}$. The GCM's ability to adapt to changing collective coordinates and its variational nature make it well-suited for describing complex states that break symmetries, allowing for a flexible and efficient description of properties such as ground states, excited states, and transitions between them.

Technically, to determine $f({\setb{\alpha}})$, substituting \eqref{gcm_wfn} into the Schr\"{o}dinger equation, and then the variational process leads to the so-called GHW integral equation:

\begin{align}
\int {\rm d}{\setb{\alpha}}' \bigg( \langle \phi(\setb{\alpha})|H|\phi(\setb{\alpha}') \rangle - E \langle \phi(\setb{\alpha})|\phi(\setb{\alpha}') \rangle \bigg) f({\setb{\alpha}}') = 0.
\label{GHW_integral}
\end{align}
This integral equation can be discretized in a finite-dimensional subspace of the many-body Hilbert space, leading to a general eigenvalue problem:
\begin{align}
     \sum_{j=1}^{N_\alpha} \bigg( \mathbf{H}_{ij} - E \mathbf{S}_{ij} \bigg) f(\alpha_j) = 0,
\end{align}
where
\begin{align}
    \mathbf{H}_{ij} &= \langle \phi(\alpha_i)|H|\phi(\alpha_j) \rangle, \notag \\
    \mathbf{S}_{ij} &= \langle \phi(\alpha_i)|\phi(\alpha_j) \rangle. \notag
\end{align}

While the GCM is a powerful method, it can be computationally demanding, especially when dealing with a large number of basis states or when symmetry-breaking effects are intricate. For strongly correlated molecular systems, these requirements are often associated with incorporating sufficient electron correlation effects into GCM calculations, which typically involves using more sophisticated electronic structure methods than mean-field theory. Techniques such as configuration interaction (CI), coupled-cluster (CC) methods, and many-body perturbation theory (MBPT) can be employed to capture electron correlation effects accurately~\cite{shavitt2009many}. The basis functions generated from these methods should be capable of representing the electron correlation effects relevant to the quantum phenomena under investigation. Various levels of electron correlation can be included in GCM calculations, depending on the chosen basis functions and computational resources available.
It is interesting to observe the GCM as a generalization of the non-orthogonal determinants (e.g., non-orthogonal configuration interaction). Even though non-entangled single-particle states were employed in the GCM for simplifying the implementation in our previous report~\cite{qugcm}, the entangled many-body states can also be employed in the framework to involve complex correlations among particles, making it suitable for describing systems with complex shapes and collective motion.


\section{Mapping optimization to a configuration subspace} \label{sec:theorem}

\noindent
\textbf{Theorem 1.} For a Hamiltonian $H$ and a parametrized normalized ansatz $|\Phi(\vect{\theta})\rangle = U(\vect{\theta})|\phi\rangle$, where $|\phi\rangle$ is a fixed reference state and $U(\vect{\theta})$ is a unitary operator dependent on the parameter $\vect{\theta} = (\theta_1,\theta_2,\cdots \theta_P)$, the minimal expectation value 
\begin{align}
\epsilon = \min_{\vect{\theta}} \langle \phi | U^\dagger(\vect{\theta})|H|U(\vect{\theta})|\phi \rangle \label{eq:expectation}  
\end{align}
is the upper bound of the minimal eigenvalue of $H$ in an arbitrary linearly independent $M$-configuration subspace $\{\psi_1,\psi_2,\cdots, \psi_M\}$ ($M\ge P$) where $U(\vect{\theta})|\phi\rangle$ can be completely projected out, i.e.,
\begin{align}
    U(\vect{\theta})|\phi\rangle = \sum_{i=1}^M c_i (\vect{\theta}) |\psi_i\rangle. \label{eq:expansion_A1},
\end{align}
where
\begin{align}
    c_i(\vect{\theta}) = \sum_{j=1}^M (\mathbf{S}^{-1})_{ij} \langle \psi_j | U(\vect{\theta}) | \phi \rangle
\end{align}
and $\mathbf{S}$ is the  the overlap matrix with
\begin{align}
    \mathbf{S}_{ij} = \langle \psi_i | \psi_j \rangle.
\end{align}

\noindent
\textbf{Proof~:} Given (\ref{eq:expansion_A1}), $U(\vect{\theta})|\phi \rangle$ is one state in the linearly independent $M$-configuration subspace, thus the minmal expectation value from minimizing (\ref{eq:expectation}) is constrained by the form of $U(\vect{\theta})$, and does not necessarily give the minimal expectation value of $H$ in the subspace. Instead, searching the minimal expectation value of $H$ in the subspace can be re-formulated to minimizing a generalized Rayleigh quotient for the projected Hamiltonian matrix $\mathbf{H}$ and the overlap matrix $\mathbf{S}$
\begin{align}
    \min_{f}~ R(f) = \frac{f^\ast \mathbf{H} f}{f^\ast \mathbf{S} f}, ~~~~\text{subject to}~~f^\ast \mathbf{S} f = 1,
\end{align}
with matrix elements defined as
\begin{align}
    \mathbf{H}_{ij} &= \langle \psi_i | H | \psi_j \rangle, \\
    \mathbf{S}_{ij} &= \langle \psi_i | \psi_j \rangle.
\end{align}
By introducing a Lagrange multiplier and forming the Lagrangian,
\begin{align}
    L(f,\epsilon) = f^\ast \mathbf{H} f - \epsilon (f^\ast \mathbf{S} f - 1),
\end{align}
the minimization corresponds to taking the derivative of $L(f,\epsilon)$ with respect to $f$ and setting it to zero, leading to a generalized eigenvalue equation:
\begin{align}
    \mathbf{H} f = \epsilon \mathbf{S} f, \label{eq:general_eigen}
\end{align}
where the lowest eigenvalue corresponds to the global minimum in the $M$-configuration subspace.\\

\noindent
\textbf{Lemma 1} Consider a unitary operation $U$ parametrized by a single scalar 
$\theta$ in a variational ans{\"a}tze. If, for any value of $\theta$, the action of $U$
on $|\phi\rangle$ produces at most two distinct configurations, then the task of minimizing $\langle \phi | U^\dagger(\vect{\theta})|H|U(\vect{\theta})|\phi \rangle$ can be equivalently approached by solving the following $2\times 2$ generalized eigenvalue problem:
\begin{align} 
&\left( \begin{array}{cc} 
\langle \phi | U^\dagger(\theta_1) H U(\theta_1)|\phi \rangle & \langle \phi | U^\dagger(\theta_1) H U(\theta_2)|\phi \rangle \\ 
\langle \phi | U^\dagger(\theta_2) H U(\theta_1)|\phi \rangle & \langle \phi | U^\dagger(\theta_2) H U(\theta_2)|\phi \rangle 
\end{array} \right) f \notag \\
&~~= \epsilon \left( \begin{array}{cc} 
1 & \langle \phi | U^\dagger(\theta_1) U(\theta_2)|\phi \rangle \\
\langle \phi | U^\dagger(\theta_2) U(\theta_1)|\phi \rangle & 1 
\end{array} \right) f. \label{2by2_eigen}
\end{align} 
Here $\theta_1$ and $\theta_2$ are chosen such that $\|\langle \phi | U^\dagger(\theta_1) U(\theta_2)|\phi \rangle\|^2 \neq 1$. The lowest eigenvalue of the $2 \times 2$ problem yields the minimal energy within the defined subspace.\\

\noindent
\textbf{Proof :} For any $\theta$, let's express
\begin{align}
    U(\theta)|\phi\rangle = a(\theta)|\phi_1\rangle + b(\theta) |\phi_2\rangle
\end{align}
where $|\phi_1\rangle$ and $|\phi_2\rangle$ are orthogonal, i.e., $\langle \phi_1 | \phi_2\rangle = 0$, and $a(\theta)$ and $b(\theta)$ are scalar functions of $\theta$ with $|a(\theta)|^2 + |b(\theta)|^2 = 1$ 

According to \textbf{Theorem 1}, the minimization of $\langle \phi | U^\dagger(\vect{\theta})|H|U(\vect{\theta})|\phi \rangle$ over all $\theta$ can be recast into a $2\times 2$ eigenvalue problem with the Hamiltonian matrix
\begin{align}
    \mathbf{H} = \left( \begin{array}{cc}
    \langle \phi_1 | H | \phi_1 \rangle & 
    \langle \phi_1 | H | \phi_2 \rangle \\ 
    \langle \phi_2 | H | \phi_1 \rangle & 
    \langle \phi_2 | H | \phi_2 \rangle    
    \end{array} \right).
\end{align}
Alternatively, the expansion of $U(\theta)|\phi\rangle$ can also be expanded by non-orthogonal basis, e.g.
\begin{align}
    U(\vect{\theta})|\phi\rangle = c_1 U(\theta_1)|\phi\rangle + c_2 U(\theta_2)|\phi\rangle,
\end{align}
where $\theta_1$ and $\theta_2$ are two scalars satisfying $\|\langle \phi | U^\dagger(\theta_1) U(\theta_2) | \phi\rangle\|^2 \neq 1$, and $c_1$ and $c_2$ depends on the choice of $\theta, \theta_1, \theta_2$ and can be obtained through
\begin{align}
    (a(\theta_1)b(\theta_2)-b(\theta_1)a(\theta_2)) c_1 &= a(\theta)b(\theta_2)-b(\theta)a(\theta_2), \\
    (a(\theta_2)b(\theta_1)-b(\theta_2)a(\theta_1)) c_2 &= a(\theta)b(\theta_1)-b(\theta)a(\theta_1).
\end{align}
If the expansion is done in such a non-orthogonal subspace, the minimization of $\langle \phi | U^\dagger(\vect{\theta})|H|U(\vect{\theta})|\phi \rangle$ transforms into a generalized eigenvalue problem (\ref{eq:general_eigen}) with matrices defined as
\begin{align}
    \mathbf{H} = \left( \begin{array}{cc}
    \langle \phi | U^\dagger(\theta_1) H U(\theta_1)|\phi \rangle & 
    \langle \phi | U^\dagger(\theta_1) H U(\theta_2)|\phi \rangle \\ 
    \langle \phi | U^\dagger(\theta_2) H U(\theta_1)|\phi \rangle & 
    \langle \phi | U^\dagger(\theta_2) H U(\theta_2)|\phi \rangle    
    \end{array} \right).
\end{align}
and
\begin{align}
    \mathbf{S} = \left( \begin{array}{cc}
    1 & \langle \phi | U^\dagger(\theta_1) U(\theta_2)|\phi \rangle \\
    \langle \phi | U^\dagger(\theta_2) U(\theta_1)|\phi \rangle & 1 
    \end{array} \right).
\end{align}
Solving this generalized eigenvalue problem allows us to determine the combination of $U(\theta_1)|\phi\rangle$ and $U(\theta_2)|\phi\rangle$ that corresponds to the lowest expectation value for $H$ in this two-configuration subspace. \\


\section{Energy gradient in ADAPT-GCIM framework} \label{sec:grad}

In the GCIM framework, the $k$-th eigenvalue of the projected $H$, $\epsilon_k$, can be represented as 
\begin{equation}
\epsilon_k = \frac{f^T_k \mathbf{H} f_k}{f^T_k \mathbf{S} f_k},
\label{eigendisc}
\end{equation}
where $f_k$ is the corresponding normalized eigenvectors, $f^T_k f_k = 1$. Assume the working subspace comprises $k$ generating functions, $\{| \psi_i \rangle \}_{1\le i\le k}$, in which $k-1$ are generated by acting one Givens rotation on the reference, and only one is generated by acting the product form of Givens rotation on the reference, i.e.
\begin{align}
|\psi_i \rangle = \left\{ \begin{array}{rc}
  G_{p_i,q_i}(\theta_i)|\phi_0\rangle,   &  1\le i \le k-1\\
  \prod_{j=1}^{k-1}G_{p_j,q_j}(\theta_j)|\phi_0\rangle,   & i = k
\end{array}  \right.   ~~.
\end{align}
Then ${\bf H}$ and ${\bf S}$ are $k\times k$ matrices projected in this working subspace with their matrix elements being defined as
\begin{align}
{\bf H}_{ij} &= \langle\psi_i|H|\psi_j\rangle , \label{H_pq} \\
{\bf S}_{ij} &= \langle\psi_i|\psi_j\rangle , \label{S_pq} 
\end{align}
and the derivative of the eigenvalue $\epsilon_k$ with respect to the rotation $\theta_s$ can be computed through
\begin{align}
\frac{\partial \epsilon_k}{\partial \theta_s} = \frac{ \langle\frac{\partial \mathbf{H}}{\partial \theta_s}\rangle_k \cdot \langle \mathbf{S} \rangle_k - \langle \mathbf{H} \rangle_k \cdot \langle \frac{\partial \mathbf{S}}{\partial \theta_s}\rangle_k}{\langle \mathbf{S} \rangle_k^2},
\end{align}
with $\langle \mathbf{O} \rangle_k = f^T_k \mathbf{O} f_k $ for matrix $\mathbf{O}$. The matrix elements of $\frac{\partial \mathbf{S}}{\partial \theta_s}$ and $\frac{\partial \mathbf{H}}{\partial \theta_s}$ are the following,
\begin{itemize}
    \item if neither $|\psi_i\rangle$ nor $|\psi_j\rangle$ is parametrized by $\theta_s$
    \begin{align}
        \frac{\partial (\mathbf{H}_{ij})}{\partial \theta_s} = \frac{\partial (\mathbf{S}_{ij})}{\partial \theta_s} = 0 ,
    \end{align}
    \item if only $|\psi_i\rangle$ is parametrized by $\theta_s$ and $i\neq j$
    \begin{align}
        \frac{\partial (\mathbf{H}_{ij})}{\partial \theta_s} &= \langle\psi_i| (a_{q_s}^\dagger a_{p_s} - a_{p_s}^\dagger a_{q_s}) H|\psi_j\rangle, \\ 
        \frac{\partial (\mathbf{S}_{ij})}{\partial \theta_s} &= \langle\psi_i| (a_{q_s}^\dagger a_{p_s} - a_{p_s}^\dagger a_{q_s}) |\psi_j\rangle, 
    \end{align}
    \item if only $|\psi_j\rangle$ is parametrized by $\theta_s$ and $i\neq j$
    \begin{align}
        \frac{\partial (\mathbf{H}_{ij})}{\partial \theta_s} &= \langle\psi_i| H (a_{p_s}^\dagger a_{q_s} - a_{q_s}^\dagger a_{p_s}) |\psi_j\rangle,   \\ 
        \frac{\partial (\mathbf{S}_{ij})}{\partial \theta_s} &= \langle\psi_i| (a_{p_s}^\dagger a_{q_s} - a_{q_s}^\dagger a_{p_s}) |\psi_j\rangle,   
    \end{align}
    \item if both $|\psi_j\rangle$ are parametrized by $\theta_s$
    \begin{align}
        \frac{\partial (\mathbf{H}_{ij})}{\partial \theta_s} &= \langle\psi_i| [H, (a_{p_s}^\dagger a_{q_s} - a_{q_s}^\dagger a_{p_s})] |\psi_j\rangle,   \label{eq:grad1}\\ 
        \frac{\partial (\mathbf{S}_{ij})}{\partial \theta_s} &= 0.  \label{eq:grad2}
    \end{align}
\end{itemize}
As can be seen matrices $\frac{\partial \mathbf{S}}{\partial \theta_s}$ and $\frac{\partial \mathbf{H}}{\partial \theta_s}$ are very sparse with the non-zero elements only residing in the column and row whose index is equal to $p$, which can greatly reduce the computational complexity. The gradient employed in the proposed ADAPT-GCIM in Figure 3 can be considered as an approximation to the above gradient calculations, where only (\ref{eq:grad1}) and (\ref{eq:grad2}) are taken into consideration.\\


\section{ADAPT-VQE-GCIM and ADAPT-VQE-GCIM(1) algorithms} \label{sec:adapt_vqe_gcm}

The main structure of ADAPT-VQE-GCIM in Algorithm~\ref{alg:gcm-vqe} is built upon a complete ADAPT-VQE process from Ref.~\citenum{Grimsley2019}. 
In fact, the essential use of ADAPT-VQE is to provide the unitary operators and phase values (except the phase for the latest selected ansatze) to form the bases for GCIM solvers. ADAPT-VQE-GCIM(1) only includes the selected single Givens rotations from all previous ADAPT iterations and the final product of the Givens rotations when ADAPT-VQE converges.

\begin{algorithm}[H]
  \caption{ADAPT-VQE-GCIM \protect\label{alg:gcm-vqe}}
   \begin{algorithmic}[1]
   \Require Operator pool $\{G_{l}\}_{(l)\in L}$ with $L$ the orbital index pool, initial state $\ket{\psi_{(0)}^{VQE}}:= \ket{\phi_{HF}}$, VQE optimal parameter vector $\star{\vect{\Theta}}_{(0)} = \emptyset$, GCIM basis set $\mathbf{B} = \emptyset$, and Hamiltonian $H$
   \State $k \leftarrow 0$
   \While{convergence criteria is not met}
        \State $k \leftarrow k+1$
        \State \begin{varwidth}[t]{\linewidth}
        \textbf{Select operator:}\par
        \hskip\algorithmicindent $G_{(k)} \leftarrow \max_{l\in L}\left\{\frac{\partial \epsilon}{\partial \theta_{l}}\big|_{\theta_l = 0} = \bra{\psi^{VQE}_{(k-1)}}\left[H, A_{l}\right] \ket{\psi^{VQE}_{(k-1)}}\right\}$;
        \end{varwidth}
        \State \begin{varwidth}[t]{\linewidth}
        \textbf{Prepare VQE ansatz and optimize free parameters:}\par
        \hskip\algorithmicindent $\ket{\tilde{\psi}_{(k)}^{VQE}(\vect{\Theta}_{(k)})} = \prod_{i=1}^k G_{(i)}(\theta_i)\ket{\phi_{HF}}$;\par
        \hskip\algorithmicindent $\star{\vect{\Theta}}_{(k)} = \argmin_{\vect{\Theta}_{(k)}} \langle \tilde{\psi}_{(k)}^{VQE}(\vect{\Theta}_{(k)}) | H \ket{\tilde{\psi}_{(k)}^{VQE}(\vect{\Theta}_{(k)})}$;\par 
        \hskip\algorithmicindent $\ket{\psi_{(k)}^{VQE}} \leftarrow \ket{\tilde{\psi}_{(k)}^{VQE}(\vect{\Theta}^\ast_{(k)})}$;
        \end{varwidth}
        \State \begin{varwidth}[t]{\linewidth}
        \textbf{Expand the basis set $B$:}\par
        \hskip\algorithmicindent $\mathbf{B} \leftarrow G_{(k)}(\theta^\ast_k)|\phi_{HF}\rangle$;\par
        \hskip\algorithmicindent $\mathbf{B} \leftarrow \ket{\psi_{(k)}^{VQE}}$;
        \end{varwidth}
        \State \begin{varwidth}[t]{\linewidth}
        \textbf{Compute matrices and solve for eigenvalues:}\par 
        \hskip\algorithmicindent $\mathbf{H} = \mathbf{B}^T H \mathbf{B}$;\par
        \hskip\algorithmicindent $\mathbf{S} = \mathbf{B}^T \mathbf{B}$;\par
        \hskip\algorithmicindent $\mathbf{H} f = \epsilon \mathbf{S} f\label{line:qugcm2}$;
        \end{varwidth}
   \EndWhile
   \State \Return eigenvalues $\epsilon$ after convergence
   \end{algorithmic}
\end{algorithm}


\section{Operator pool construction and implementation} \label{sec:implement}

\subsection{Operator pool construction}

\textcolor{black}{
We use the parameter-efficient operator pool proposed in Ref.~\citenum{tang2021qubit} in the numerical experiments to reduce the number of ADAPT iterations for ADAPT-VQE and ADAPT-GCIM. Such pool consists of spin-adapted single-excitation operators
\begin{align}
    T^i_{p,q} \propto \left(a^\dagger_{p_{\uparrow,i}} a_{q_{\uparrow,i}} + a^\dagger_{p_{\downarrow,i}} a_{q_{\downarrow,i}} \right) - h.c. \label{eq:op-single}
\end{align}
and spin-adapted double-excitation operators, which include singlets
\begin{align}
    T^{S,i}_{p,q,r,s} &\propto \Big(
    a^\dagger_{p_{\uparrow,i}}a^\dagger_{q_{\downarrow,i}}a_{r_{\uparrow,i}}a_{s_{\downarrow,i}} -
    a^\dagger_{p_{\uparrow,i}}a^\dagger_{q_{\downarrow,i}}a_{r_{\downarrow,i}}a_{s_{\uparrow,i}} \nonumber\\
    &\quad- a^\dagger_{p_{\downarrow,i}}a^\dagger_{q_{\uparrow,i}}a_{r_{\uparrow,i}}a_{s_{\downarrow,i}} +
    a^\dagger_{p_{\downarrow,i}}a^\dagger_{q_{\uparrow,i}}a_{r_{\downarrow,i}}a_{s_{\uparrow,i}} 
    \Big) - h.c. \label{eq:op-double-singlet}
\end{align}
and triplets
\begin{align}
    T^{T,i}_{p,q,r,s} &\propto \Big(
    2a^\dagger_{p_{\uparrow,i}}a^\dagger_{q_{\uparrow,i}}a_{r_{\uparrow,i}}a_{s_{\uparrow,i}} +
    a^\dagger_{p_{\uparrow,i}}a^\dagger_{q_{\downarrow,i}}a_{r_{\uparrow,i}}a_{s_{\downarrow,i}} \nonumber \\
    &\quad+ a^\dagger_{p_{\uparrow,i}}a^\dagger_{q_{\downarrow,i}}a_{r_{\downarrow,i}}a_{s_{\uparrow,i}} +
    a^\dagger_{p_{\downarrow,i}}a^\dagger_{q_{\uparrow,i}}a_{r_{\uparrow,i}}a_{s_{\downarrow,i}} \nonumber\\
    &\quad+ a^\dagger_{p_{\downarrow,i}}a^\dagger_{q_{\uparrow,i}}a_{r_{\downarrow,i}}a_{s_{\uparrow,i}} +
    2a^\dagger_{p_{\downarrow,i}}a^\dagger_{q_{\downarrow,i}}a_{r_{\downarrow,i}}a_{s_{\downarrow,i}}
    \Big) - h.c., \label{eq:op-double-triplet}
\end{align}
where $p,q,r,s$ are spatial orbitals used in Operator $i$,  $\uparrow$ is spin up, $\downarrow$ is spin down, and $h.c.$ stands for the Hermitian conjugate (of all previous terms). To simplify expressions, we use the mapping 
\begin{align*}
    g_{\uparrow,i}  = 2g  \text{ and } g_{\downarrow,i}  = 2g+1
\end{align*}
to re-index all spin orbitals in Operator $i$ and denote a spin orbital as $g_i$, for $g \in \{p,q,r,s\}$. All orbital indices start from 0.
We illustrate first 11 selected operators by ADAPT-GCIM in the $H_4$ (linear) experiments in Table~\ref{tab:lh4ansatz}. They are all double-excitation operators, while the one in the Iteration 10 is a triplet and the rest are singlets.
}

\textcolor{black}{Each skew-Hermitian pair of fermionic operators in \eqref{eq:op-single}-\eqref{eq:op-double-triplet}, i.e., $a^\dagger_{p_i}a_{q_i}-h.c.$ or $a^\dagger_{p_i}a^\dagger_{q_i}a_{r_i}a_{s_i}-h.c.$, can be translated into a linear combination of Pauli strings under Jordan-Wigner mapping, as shown in the next section.}

\begin{table}
\textcolor{black}{
    \centering
    \caption{Selected operators by ADAPT-GCIM in the $H_4$ (linear) experiment, $[p_i^\dagger, q_i^\dagger, r_i, s_i] \rightarrow a^\dagger_{p_i}a^\dagger_{q_i}a_{r_i}a_{s_i}$.}
    \label{tab:lh4ansatz}
    \begin{tabular}{ccl} \hline \hline
        \multirow{2}{*}{Iter.} &$p,q,r,s$   &Operator, omit normalization coeff. \\
          &(Spatial Orbitals)   &(Spin Orbitals) \\ \hline
        \multirow{2}{*}{1}  &\multirow{2}{*}{$0,1,2,3$} &   $[2^\dagger,1^\dagger,6, 5]-[2^\dagger,1^\dagger,7, 4]$ \\
           & &$\,-[3^\dagger,0^\dagger,6, 5]+[3^\dagger,0^\dagger,7, 4] -h.c. $  \\
        2 &$1,1,2,2$  &    $[3^\dagger,2^\dagger,5, 4] -h.c.$ \\
        3  &$1,1,3,3 $   &  $[3^\dagger,2^\dagger,7, 6] -h.c.$  \\
        4 &$0,0,1,1 $   & $[1^\dagger,0^\dagger,3, 2] -h.c.$\\
        5 &$0,0,3,3$    & $[1^\dagger,0^\dagger,7, 6]-h.c.$\\
        6 &$0,0,2,2 $   & $[1^\dagger,0^\dagger,5, 4] -h.c.$\\
        7 &$2,2,3,3 $   & $[5^\dagger,4^\dagger,7, 6]-h.c.$\\
        8 &$1,3,2,2 $   & $[5^\dagger,4^\dagger,6, 3] - [5^\dagger,4^\dagger,7, 2]-h.c.$\\
        9 &$1,3,3,3 $   & $-[6^\dagger,3^\dagger,7, 6]+[7^\dagger,2^\dagger,7, 6] -h.c.$\\
        \multirow{3}{*}{10} &\multirow{3}{*}{$0,1,0,3 $}   & $2 \cdot [2^\dagger,0^\dagger,6, 0] + [2^\dagger,1^\dagger,6, 1]$  \\
        & & $\,+ [2^\dagger,1^\dagger,7, 0] + [3^\dagger,0^\dagger,6, 1]$  \\
        & & $\,+ [3^\dagger,0^\dagger,7, 0] + 2\cdot [3^\dagger,1^\dagger,7, 1]-h.c.$\\
        11 &$0,2,3,3 $   & $ -[4^\dagger,1^\dagger,7, 6]+ [5^\dagger,0^\dagger,7, 6]-h.c.$\\
     \hline \hline
    \end{tabular}
    }
\end{table}

\subsection{Operator implementation and resource estimation} 

For skew-Hermitian operators 
\begin{align}
    A_{p_i,q_i} &= a_{p_i}^\dagger a_{q_i} - a_{q_i}^\dagger  a_{p_i},~~q_i < p_i \label{eq:skew_single} \\
    A_{p_i,r_i,q_i,s_i} &= a_{p_i}^\dagger a_{r_i}^\dagger a_{q_i} a_{s_i} - a_{s_i}^\dagger a_{q_i}^\dagger a_{r_i} a_{p_i}, ~~q_i < s_i < p_i < r_i \label{eq:skew_double} 
\end{align}
their evolution in the qubit space can be expressed as
\begin{align}
    &\exp(\theta A_{p_i,q_i}) = \exp\bigg( - {\rm i} \frac{\theta}{2}
    (X_{q_i}Y_{p_i} - Y_{p_i}X_{q_i}) \prod_{l=q_i+1}^{p_i-1} Z_l
    \bigg), \label{eq:app-exp-single}\\
    &\exp(\theta A_{p_i,r_i,q_i,s_i}) = \exp\bigg( - {\rm i} \frac{\theta}{8}
    (X_{q_i}Y_{s_i}X_{p_i}X_{r_i} + Y_{q_i}X_{s_i}X_{p_i}X_{r_i} \notag \\
    &~~+Y_{q_i}Y_{s_i}Y_{p_i}X_{r_i} + Y_{q_i}Y_{s_i}X_{p_i}Y_{r_i} - X_{q_i}X_{s_i}Y_{p_i}X_{r_i} - X_{q_i}X_{s_i}X_{p_i}Y_{r_i} \notag \\
    &~~ -Y_{q_i}X_{s_i}Y_{p_i}Y_{r_i} - X_{q_i}Y_{s_i}Y_{p_i}Y_{r_i}) \prod_{k=p_i+1}^{r_i-1} Z_k \prod_{l=q_i+1}^{s_i-1} Z_l
    \bigg),\label{eq:app-exp-double}
\end{align}
under the Jordan-Widger mapping
\begin{align}
    a_{p_i} = \frac{1}{2} (X_{p_i} + {\rm i} Y_{p_i}) \prod_{k=0}^{p_i-1} Z_k,~~~
    a_{p_i}^\dagger = \frac{1}{2} (X_{p_i} - {\rm i} Y_{p_i}) \prod_{k=0}^{p_i-1} Z_k.
\end{align}
\textcolor{black}{Thus, the evolution of operators \eqref{eq:op-single}-\eqref{eq:op-double-triplet} can be trotterized as the products of \eqref{eq:app-exp-single} and \eqref{eq:app-exp-double}, respectively.}

Next, a common way to implement unitary operators $\exp(\theta A_{p_i,q_i})$ and $\exp(\theta A_{p_i,r_i,q_i,s_i})$ in quantum circuits is to trotterize them into the product of the exponentials of single Pauli strings~\cite{nielsen2010quantum}. Without any circuit optimization, the standard method has the CNOT gate counts $4(p_i - q_i)$ for \eqref{eq:app-exp-single} and $16(s_i - q_i + r_i - p_i +1)$ for \eqref{eq:app-exp-double}~\cite{barkoutsos2018quantum}. In Ref.~\citenum{PhysRevA.102.062612}, researchers reduce the CNOT gate counts to $2(p_i - q_i)+1$ and $2(s_i - q_i +r_i - p_i) + 9$, respectively. To align with the quantum resource estimation for ADAPT-VQE in Ref.~\citenum{tang2021qubit}, we use the standard way to obtain the data in Figure 5. While the absolute numbers will drop significantly if we use the advanced implementations, the relative difference between ADAPT-VQE and ADAPT-GCIM still holds because both algorithms use exactly the same operator pool for each geometry in the simulation.

Alternatively, excitation operators can be directly implemented in terms of Givens rotation operators in circuits. For single excitations, Eq. (\ref{eq:skew_single}), its corresponding Givens rotation operator for two adjacent indices is given in Figure 1, i.e., two CNOT gates. If the two indices are not adjacent, $p_i - q_i - 1$ FSWAP gates will be employed on each side of the Givens rotation operator. If each FSWAP contains three CNOT gates,~\cite{hashim2021optimized} the total number of CNOT gates for implementing a single excitation Givens rotation would be $6(p_i - q_i) - 4$. For double excitations, Eq. (\ref{eq:skew_double}), a 4-qubit Givens rotation requires 14 CNOT gates assuming $p_i, r_i, q_i, s_i$ are adjacent to each other~\cite{arrazola2022universal}. 
Otherwise, $2(r_i - s_i + p_i -q_i - 4)$ FSWAP ladders will be employed, which leads to $6(r_i - s_i + p_i -q_i) -10$ CNOT gates for a double excitation Givens rotation.
In Ref. \citenum{Anselmetti_2021}, an efficient implementation for the general Givens rotation is discussed, where in the 4-qubit Hilbert space the  Givens rotation applied to two adjacent spatial orbitals that preserves the number of $\alpha$ and $\beta$ spins, and the spin-squared operator $\hat{S}^2$ can be implemented using only 4 CNOT gates.


\section{Mitigation to the numerical instability}\label{sec:orth}

In the computation of $\mathbf{H} f = \epsilon \mathbf{S} f$ within GCIM framework, the overlap matrix $\mathbf{S}$ may be indefinite in experiments, especially when affected by a finite number of shots. To address this issue, one simple method is to add a small disturbance, such as $10^{-12}$, on all diagonal entries in $\mathbf{S}$. In our simulations, we adopted a more stable solution from Ref.~\citenum{urbanek2020chemistry}. Since setting $\mathbf{S}_{ij} = \overline{\mathbf{S}}_{ji}$ ensures $\mathbf{S}$ is always Hermitian, the eigen-decomposition of $\mathbf{S}$ gives
$\mathbf{S}  = UDU^{\dagger}$
where $U$ is a unitary matrix composed of the eigenvectors of $\mathbf{S}$, and the $D$ is a diagonal matrix consisting of eigenvalues of $\mathbf{S}$. Not all diagonal entries of $D$ are positive in practice, so we only keep largest $L$ number of positive eigenvalues and truncate both $U$ and $D$ from $M \times M$ to $M \times L$ and $L \times L$, respectively. Then, the projection of $\mathbf{H}$ and $\mathbf{S}$ onto the subspace of the eigenspace of $\mathbf{S}$ is computed as following:
\begin{align}
    \mathbf{H}_{trunc}  &= U^{\dagger}_{trunc} \mathbf{H} U_{trunc} \\
    \mathbf{S}_{trunc}  &= U^{\dagger}_{trunc} \mathbf{S} U_{trunc} = D_{trunc}
\end{align}
where $U_{trunc}$ and $D_{trunc}$ are the truncated matrices $U$ and $D$, respectively. Therefore, the reformulated generalized eigenvalue problem
\begin{align}
    \mathbf{H}_{trunc} f_{trunc} = \epsilon_{trunc} D_{trunc} f_{trunc} \label{eq:new-eig-prob}
\end{align}
replaces the original one.

\begin{figure}[b]
    \includegraphics[width=\linewidth]{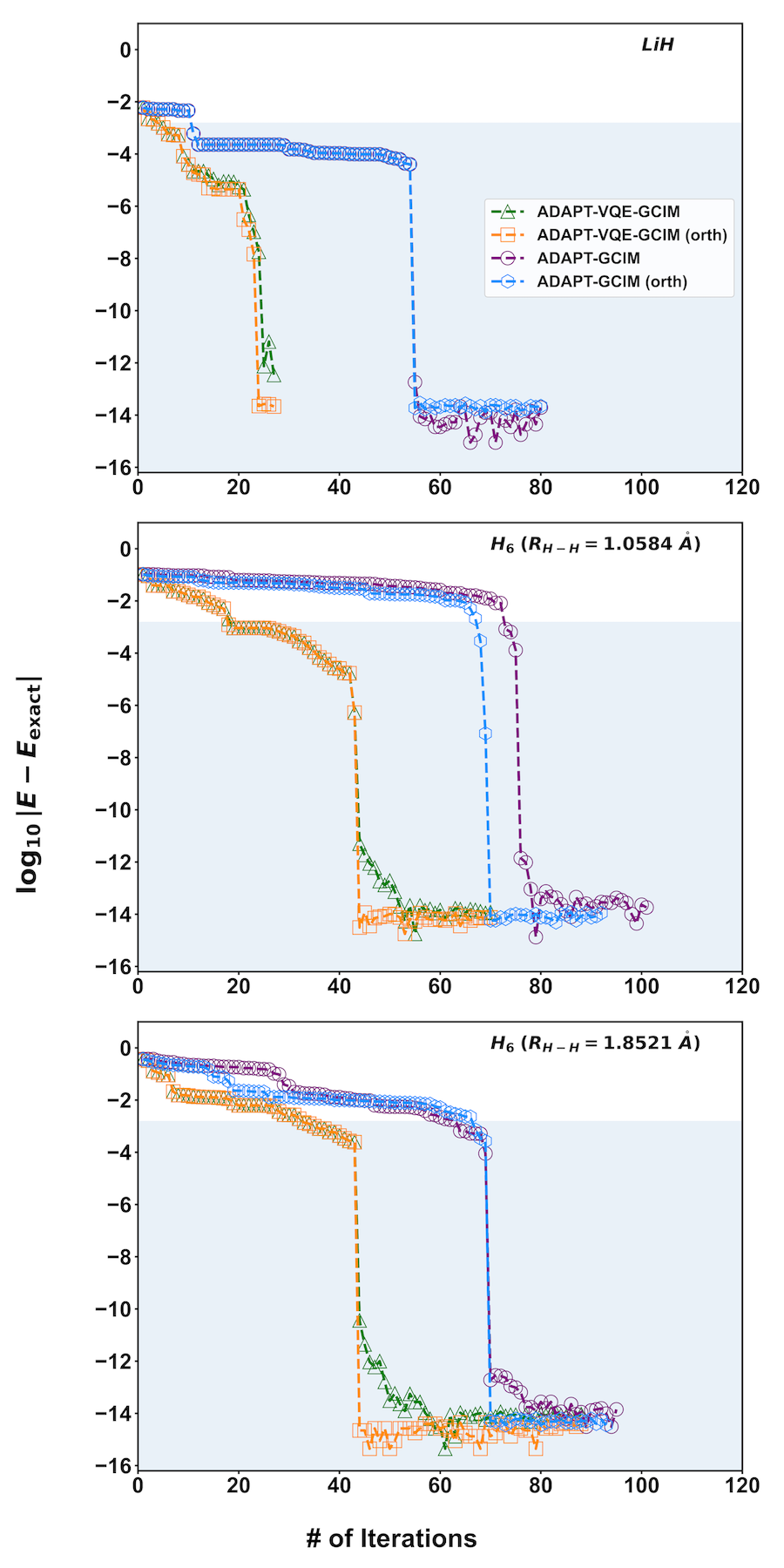}
    \caption{Convergence performance before and after classical orthogonalization of the basis sets in ADAPT-VQE-GCIM and ADAPT-VQE for selected molecules. The ``orth'' in the labels mean the GCIM calculation was done with the ``orthogonalized bases.''}
    \label{fig:orth-basis}
\end{figure}

To illustrate the effects of the quantum orthogonalization of the basis vectors, numerical simulations with classical orthogonalization on basis vectors in $LiH$ and both $H_6$ examples are conducted and demonstrated in Figure~\ref{fig:orth-basis}. It is noteworthy that there is no significant difference in the final accuracy. Combining this with the analysis in Section~\ref{sec:shot}, this subspace projection method could be effective enough to mitigate the potential numerical instability in near-future experiments.

\section{Randomness due to finite number of shots} \label{sec:shot}

\begin{figure*}[!htbp]
    \centering
    \includegraphics[width=\linewidth]{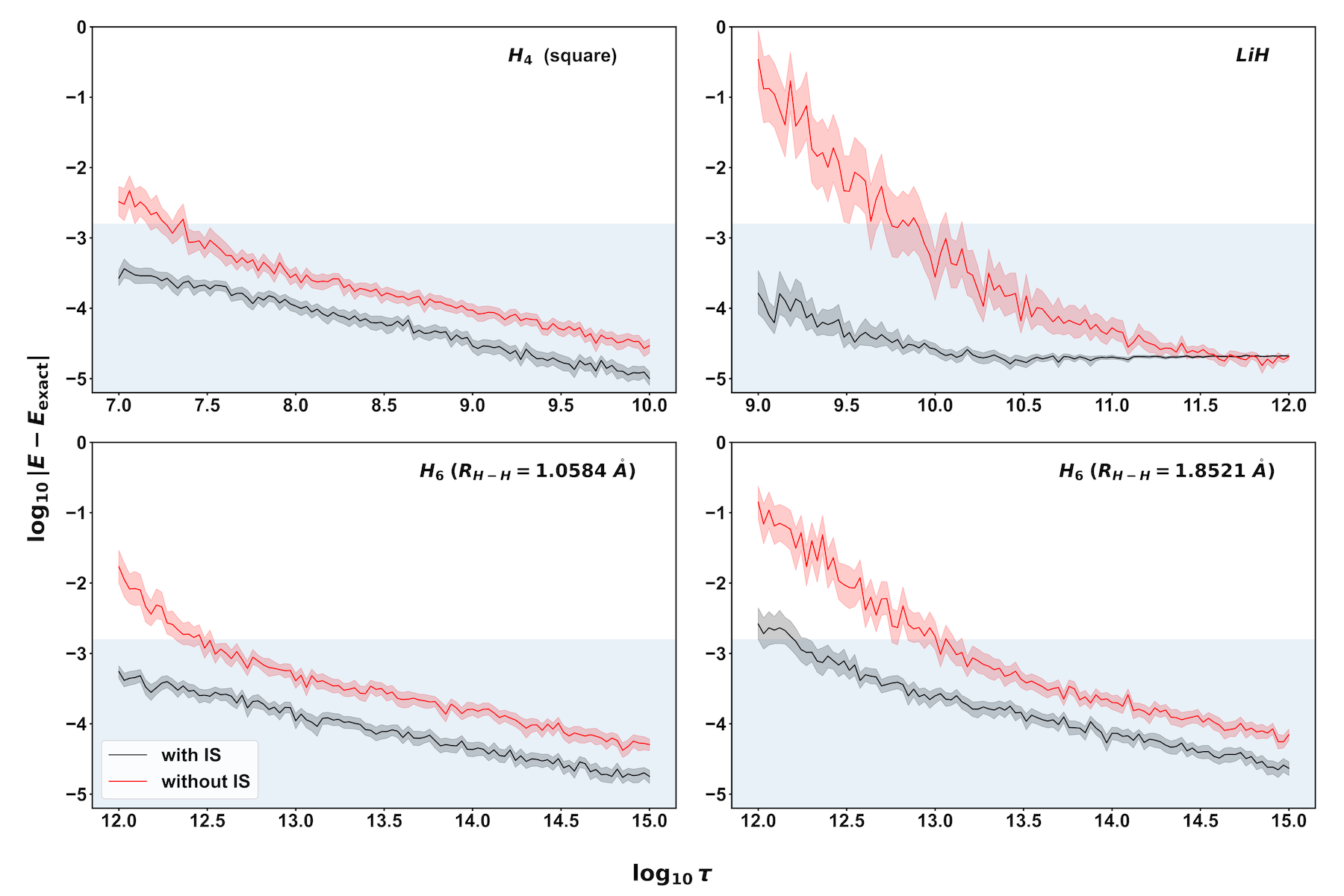}
    \caption{Monte Carlo simulations for the effects of the finite-sampling noise on the accuracy of the selected iterations in ADAPT-GCIM on different molecules. The black and red-shaded areas are the $95\%$ confidence intervals estimated from 100 runs for each given $\tau$. When not using importance sampling, $N^{k}_{shot} = \tau$ for a projected Hamiltonian $\mathbf{H}$ and $N^{k}_{shot} = 100\tau$ for an overlap matrix $\mathbf{S}$. With importance sampling, $N^{k}_{shot}$ follows~\eqref{eq:is-shot} for $\mathbf{H}$ and 100 times larger for $\mathbf{S}$. Note that, given the same value of $\tau$, the total number of shots for estimating a matrix entry is the same with or without importance sampling..
    \label{fig:shot-noise} }
\end{figure*}

Recall that in both ADAPT-GCIM and GCIM part of the ADAPT-VQE-GCIM, we compute project Hamiltonian, $\mathbf{H}$, and overlap matrix, $\mathbf{S}$, entry by entry
\begin{align}
    \mathbf{H}_{ij} &= \langle \psi_i | H | \psi_j \rangle = \langle \phi_{HF} | B_i^\dagger H B_j | \phi_{HF} \rangle \\
    \mathbf{S}_{ij} &= \langle \psi_i | \psi_j \rangle = \langle \phi_{HF} | B_i^\dagger B_j | \phi_{HF} \rangle,
\end{align}
where $B_i$ and $B_j$ are short-hand notations for some products of Givens rotation matrices for simplicity. These expectations can be computed by converting $B_i^\dagger H B_j$ and $B_i^\dagger B_j$ to the sums of Pauli strings, respectively, then grouping the Pauli strings to do simultaneous measurements. The quantum and classical complexities and effects of finite-sampling noise have been explicitly discussed in Ref.~\citenum{qugcm}. 

Instead, in this section, we illustrate the finite-sampling noise calculation when employing Hadamard test for the expectation computation. Due to the randomness of the collapse of qubits, we will transfer $\mathbf{H}_{ij}$ to a random variable where the number of shots controls its variance.

Assume the Pauli decomposition of Hamiltonian $H = \sum_{k = 1}^{N_{term}} c_kP_k$, $c_k \in \mathbb{R}$ for all $k$, has been provided, we have 
\begin{align}
    \mathbf{H}_{ij} =  \sum_{k = 1}^{N_{term}} c_k \langle \phi_{HF} | B_i^\dagger P_k B_j | \phi_{HF} \rangle. \label{eq:shot-hpauli}
\end{align}
Without the loss of the generality, we only focus on $\mathbf{H}_{ij}$ since the calculation of $\mathbf{S}_{ij}$ can be regarded as a special case for~\eqref{eq:shot-hpauli} when $N_{term} =1$, $c_k = 1$, and $P_k$ is an identity matrix. It is known that, Hadamard test gives
\begin{align}
    \Pr{\text{ancillary is } \ket{0}} &= \frac{1}{2} + \frac{1}{2} \text{Re}\langle \phi_{HF} | B_i^\dagger P_k B_j | \phi_{HF} \rangle \\
    \Pr{\text{ancillary is } \ket{1}} &= \frac{1}{2} - \frac{1}{2} \text{Re}\langle \phi_{HF} | B_i^\dagger P_k B_j | \phi_{HF} \rangle,
\end{align}
after measuring the ancillary qubit in the circuit. Thus, a binary random variable can be defined by
\begin{align}
    \Gamma^{(k)}_l = \begin{cases}
        1 & \text{if ancillary is $\ket{0}$}  \\
        -1 & \text{if ancillary is $\ket{1}$}
    \end{cases}
\end{align}
in the $l^{th}$ shot when computing $\langle B_i^\dagger P_k B_j\rangle$ with
\begin{align}
    p_k &:=  \E{\Gamma^{(k)}_l} = \text{Re}\langle \phi_{HF} | B_i^\dagger P_k B_j | \phi_{HF} \rangle, \\
    \Var{\Gamma^{(k)}_l} &= \E{\left(\Gamma^{(k)}_l\right)^2} - \left(\E{\Gamma^{(k)}_l}\right)^2 = 1 - p_k^2,
\end{align}
for all $l$ by assuming $\Gamma^{(k)}_l$'s are mutually independent and identically distributed. For simplicity, we omit the indices $i$ and $j$ for now. In other words, each measurement on the ancillary qubit in the Hadamard test circuit is modelled by sampling the random variable $\Gamma^{(k)}_l$.

We further construct another random variable, omitting indices $i$ and $j$ again,
\begin{align}
    \Lambda^{(k)} = \sum_{l = 1}^{N^{(k)}_{shot}} \Gamma^{(k)}_l
\end{align}
where $N^{(k)}_{shot}$ is the number of shots for the Hadamard test circuit of $\langle B_i^\dagger P_k B_j\rangle$, and it is considered as a user-defined constant. Then, it follows that
\begin{align}
   \frac{1}{2} \left(\Lambda^{(k)} + N^{(k)}_{shot} \right) \sim \text{Bin}\left(N^{(k)}_{shot}, \frac{1}{2}+\frac{1}{2}p_k\right),\label{eq:shot-bin}
\end{align}
by realizing $\frac{1}{2}+\frac{1}{2}\Gamma^{(k)}_i$ is a standard Bernoulli random variable.
Further combining ~\eqref{eq:shot-hpauli} and~\eqref{eq:shot-bin}, we obtain another discrete random variable 
\begin{align}
    \Xi^{(ij)} = \sum_{k = 1}^{N_{term}} \frac{c_k}{N^{(k)}_{shot}} \Lambda^{(k)} =  \sum_{k = 1}^{N_{term}} \frac{c_k}{N^{(k)}_{shot}} \sum_{l = 1}^{N^{(k)}_{shot}} \Gamma^{(k)}_l
\end{align}
that satisfies
\begin{align}
    \E{\Xi^{(ij)}} &= \sum_{k = 1}^{N_{term}}  \frac{c_k}{N^{(k)}_{shot}} N^{(k)}_{shot} p_k = \mathbf{H}_{ij} \label{eq:shot-mean}\\
    \Var{\Xi^{(ij)}} &= \sum_{k = 1}^{N_{term}} \left(\frac{c_k}{N^{(k)}_{shot}}\right)^2 N^{(k)}_{shot} \Var{\Gamma^{(k)}_l} \nonumber \\
    &= \sum_{k = 1}^{N_{term}} \frac{c^2_k}{N^{(k)}_{shot}} \left(1-p_k^2\right).  \label{eq:shot-var}
\end{align}
In this case, $\Xi^{(ij)}$ is an unbiased estimator of $\mathbf{H}_{ij}$ where the randomness is caused by the uncertainty from the collapse of the qubits during the measurement. The role of the number of shots, as~\eqref{eq:shot-var} implies, is to govern the scale of the variance. The variance~\eqref{eq:shot-var} goes to 0 when every $N_{shot}^{(k)}$ goes to infinity. 

Chebyshev's inequality shows that, for every $a>0$,
\begin{align}
    \Pr{\left| \Xi^{(ij)} - \mathbf{H}_{ij} \right| \geq a} \leq \frac{\Var{\Xi^{(ij)}}}{a^2} =: \eta_{ij}.
\end{align}
Setting $N^{(k)}_{shot}$ to $N_{shot}$ for all $i,j,k$, Table~\ref{tab:entry-diff} provides the relation between $N_{shot}$ and $ 
 \max_{i,j} \, \eta_{ij}$ when $a$ is fixed at $10^{-4}$. However, the tightness of the bound is still unclear for this discrete variable, and this is not sufficient to illustrate the effects of the disturbance of the matrix entries on the eigenvalues. This ambiguity leads to numerical simulations.

\begin{table}
    \centering
    \caption{$N_{shot}$ for the various choices of $\max_{i,j} \, \eta_{ij}$ when $a = 10^{-4}$.}
    \label{tab:entry-diff}
    \begin{tabular}{ccc} \hline \hline
        Molecule & $\max \, \eta_{ij} = 5\%$ & $\max \, \eta_{ij} = 1\%$ \\ \hline
        $H4$ (square) & $6.32\times 10^{10}$ & $3.16\times 10^{11}$ \\
        $LiH$  & $1.34\times 10^{11}$ & $6.68\times 10^{11}$ \\
        $H_6$ ($1.0584 \,\, \mathring{A})$ & $1.26\times 10^{11}$ & $6.29\times 10^{11}$ \\
        $H_6$ ($1.8521 \,\, \mathring{A})$ & $6.17\times 10^{10}$ & $3.08\times 10^{11}$ \\ \hline \hline
    \end{tabular}
\end{table}

Since it is very common to have huge $N^{(k)}_{shot}$ in practice, we can assume $\Lambda^{(k)}$ is closely approximated by
\begin{align}
    \Lambda^{(k)}_{Gauss} \sim \normal{N^{(k)}_{shot}p_k,\,\, N^{(k)}_{shot}\left(1-p_k^2\right)}
\end{align}
and mutually independent for different $k$. Thus, $\Xi^{(ij)}$ is also approximated by a Gaussian random variable $\Xi^{(ij)}_{Gauss}$ with the same mean and variance as in~\eqref{eq:shot-mean} and~\eqref{eq:shot-var}, respectively. 

The idea of importance sampling (IS) can be applied here to reduce the volatility without increasing the number of shots. By fixing the total number of shots for measuring all $P_k$'s as $\tau N_{term}$ with some large constant $\tau$, each individual Pauli string $P_k$ is allocated with  
\begin{align}
    N^{(k)}_{shot} :=  \frac{|c_k|}{\sum_{k = 1}^{N_{term}}|c_k|} \tau N_{term} \label{eq:is-shot}
\end{align}
number of shots. Under this setting, the random variable 
\begin{align}
    \Xi^{(ij)}_{IS} = \frac{\sum_{k = 1}^{N_{term}}|c_k|}{N_{term}} \cdot \frac{1}{\tau} \sum_{k = 1}^{N_{term}} \text{sign}(c_k) \Lambda^{(k)}_{Gauss} 
\end{align}
is also Gaussian and unbiased with the mean $\mathbf{H}_{ij}$ and the variance 
\begin{align}
    \Var{\Xi^{(ij)}_{IS}} = \frac{\sum_{k = 1}^{N_{term}}|c_k|}{N_{term}} \cdot \frac{1}{\tau}\sum_{k = 1}^{N_{term}} |c_k|\left(1-p_k^2\right) \label{eq:shot-var-IS}.
\end{align}
It is suggested to use larger $N^{k}_{shot}$ and $\tau$ for the overlap matrix $\mathbf{S}$ than $\mathbf{H}$ to improve the numerical stability. In the following simulations, $\mathbf{S}$ will always have 100 times larger $N^{k}_{shot}$ than $\mathbf{H}$.

The Monte Carlo simulations for squared $H_4$, $LiH$, and two $H_6$ geometries are conducted by treating each matrix entry as a normally distributed variable, following equations~\eqref{eq:shot-mean} and~\eqref{eq:shot-var}. We select a pair of $\mathbf{H}$ and $\mathbf{S}$ from a specific iteration of ADAPT-GCIM for each molecule, ensuring that the error in ground-state energy estimation is around $10^{-13}$ a.u. under the condition of an infinite number of shots. Although finite sampling gives challenges in reaching this level of accuracy, an accuracy of $10^{-4}$ to $10^{-5}$ a.u. is still attainable with the aid of importance sampling, as demonstrated in Figure~\ref{fig:shot-noise}.

\section{Intermittent and truncated optimization on ADAPT-GCIM}\label{sec:opt-gcm}

\begin{figure}[!htbp]
    \centering
    \includegraphics[width=0.99\linewidth]{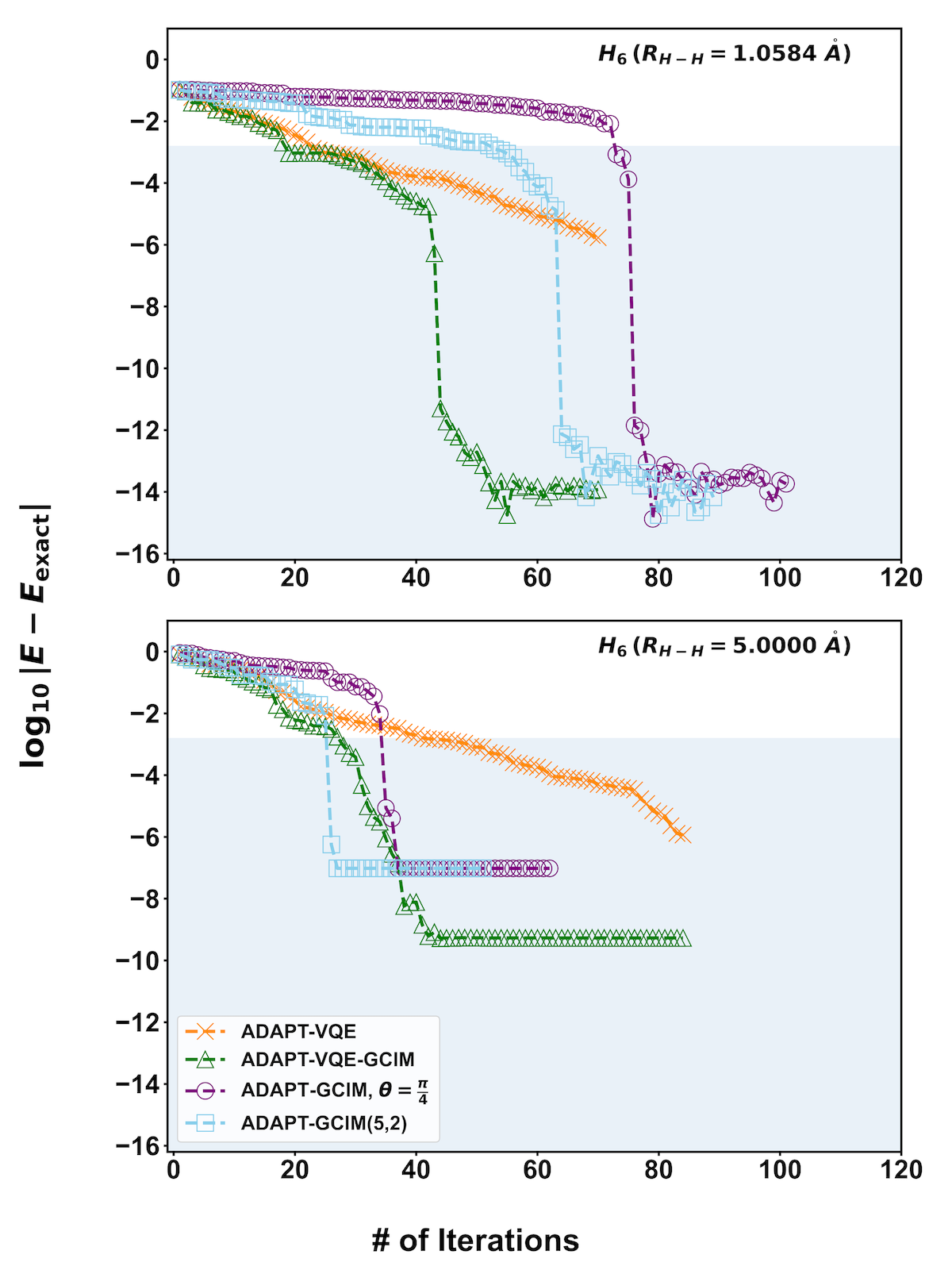}
    \caption{Convergence performance of ADAPT-GCIM with partially optimized parameters in computing the ground states. The label ``ADAPT-GCIM(5,2)'' means a parameter optimization is conducted for at most $2$ rounds in every $5$ ADAPT iterations.}
    \label{fig:truncate-exp}
\end{figure}

A middle ground between ADAPT-VQE-GCIM and ADAPT-GCIM can be drawn by allowing intermittent and truncated optimization. That is, for every $x$ ADAPT iterations (``intermittent''), at most $y$ rounds (``truncated'') of classical optimization can be conducted on ans\"{a}tze parameters over the corresponding VQE objective function. In other ADAPT iterations, the parameters for the newly added ans\"{a}ze are still initialized by a constant, such as $\pi/4$ used in the paper. 
Figure~\ref{fig:truncate-exp} illustrate the influence of such intermittent and truncated optimization on the convergence performance of ADAPT-GCIM for two geometries of $H_6$.
In the experiment, we optimized parameters for $2$ rounds in every $5$ ADAPT iterations, and denoted such method as ``ADAPT-GCIM(5,2).'' It clear to see the intermittent and truncated optimization provides a significant improvement on the convergence speed: ADAPT-GCIM(5,2) reaches $10^{-6}$ error level faster than ADAPT-VQE in both geometries. Recall that ADAPT-VQE conducted in total 4792 rounds of optimization in $R_{H-H} =1.0584 \,\, \mathring{A}$ case and 11313 rounds in $R_{H-H} = 5.0000\,\, \mathring{A}$ case, the optimization rounds in ADAPT-GCIM(5,2) were very minimal (Table IV).

\section{GCIM experiment on IBM quantum computer}\label{sec:real-exps}

\begin{figure}[!htbp]
    \centering
    \includegraphics[width=0.99\linewidth]{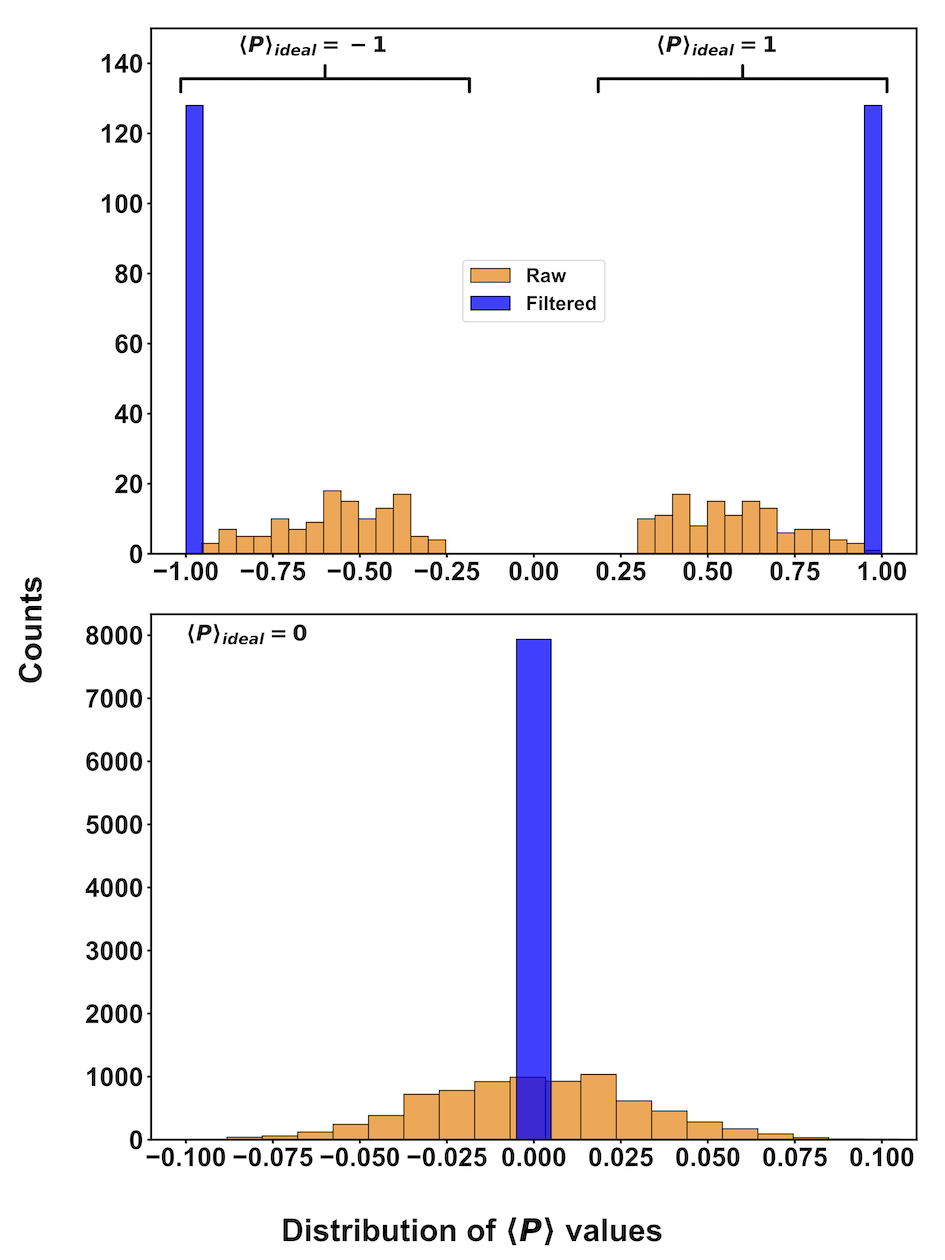}
    \caption{Distribution of the expectations of 8192 unique Pauli strings, measured from \textit{ibm\_osaka} before and after the filtering in~\eqref{eq:real-filter}. Note that, raw data is already processed by the built-in measurement error mitigation and ZNE in Qiskit Runtime.}
    \label{fig:real-filter}
\end{figure}

Employing classical preprocessing and the expectation estimator in Qiskit Runtime~\cite{Qiskit}, we evaluated the ground-state energy based on GCIM-circuit measurements from an IBM superconducting quantum computer, $ibm\_osaka$. The test molecule was the strongly correlated linear $H_4$ ($R_{H-H} = 5.0000 \,\, \mathring{A}$). Through simulation in a noise-free environment, ADAPT-GCIM reached $3.9 \times 10^{-9}$ error on ground-state energy estimation (including the Trotterization error) with 30 bases. Thus, we needed to evaluate $\frac{1}{2}(30^2 + 30) = 465$ entries for $\mathbf{H}$ and $\mathbf{S}$, respectively. In our previous work~\citenum{qugcm}, we proposed a constant-depth circuit after heavy classical preprocessing. That is, during the computation
\begin{align*}
    \mathbf{H}_{ij} = \bra{\psi_i} H \ket{\psi_j} = \bra{\phi_{HF}} U^{\dagger}_i(\theta) H U_j(\theta) \ket{\phi_{HF}}, 
\end{align*}
where $\theta = \pi/4$ as in other experiments, $ \ket{\phi_{HF}}$ is the HF state, we classically decompose the observable $U^{\dagger}_j(\pi/4) H U_i(\pi/4)$ into the linear combination of Pauli strings $\sum_k P_k$. So, quantum computer's task was to evaluate a large number of $\bra{\phi_{HF}}P_k\ket{\phi_{HF}}$. By excluding the repeated Pauli strings across 465 entries, there are 8192 unique Pauli strings (8192 circuits) need to compute in this experiment. Of course, since $\bra{\phi_{HF}}P_k\ket{\phi_{HF}} = 0$ for any $P_k$ that contains Pauli-$X$ or Pauli-$Y$, the quantum resources required can be further reduced by examining every elements in all 8192 Pauli strings. in exchange for classical running time.

We measured 1024 shots for each of 8192 circuits. With Qiskit Runtime's default measurement error mitigation and zero-noise extrapolation (ZNE)~\cite{temme2017error}, the final error on ground-state energy estimation is $0.046$. Considering $\bra{\phi_{HF}}P\ket{\phi_{HF}} \in \{-1,0,1\}$ for any Pauli string $P$, we can efficiently build a filter to further mitigate quantum errors
\begin{align}
    \bra{\phi_{HF}}P\ket{\phi_{HF}} = \begin{cases}
        1, &\text{if the noisy estimation $>0.2$}\\
        -1 &\text{if the noisy estimation $<-0.2$} \\
        0, &\text{otherwise}
    \end{cases} \label{eq:real-filter}
\end{align}
where the threshold $0.2$ is arbitrarily selected. Figure~\ref{fig:real-filter} clearly illustrate the distribution of 8192 expectations and the results after this filtering. With this problem-specific error mitigation method, error on ground-state energy estimation is reduced to $3.9 \times 10^{-9}$ from the original $0.046$. In other words, we filtered out all the quantum errors and obtained the same result as in the classical simulation.

The purpose of the experiment is to examine the possibility of GCIM framework on the current quantum machines, although the heavy classical preprocessing undermines the potential quantum advantages. We will explore this possibility further with a more entangled circuit design based on Ref.~\citenum{tang2021qubit, berry2007efficient, Anselmetti_2021, arrazola2022universal, huggins2020non, cortes2022quantum, motta2024subspace, marti2024spin, kottmann2023molecular, kottmann2024quantum, burton2024accurate}.

\bibliography{ref}

\end{document}